\documentclass[a4paper,12pt]{article}
\pdfoutput=1
\usepackage{feynmp-auto,expdlist,setspace,etoolbox}
\usepackage{amsmath, amsfonts, amssymb,mathbbol}
\usepackage{graphicx}
\usepackage{enumerate,cancel}
\usepackage{hyperref}
\usepackage{latexsym}
\usepackage{dsfont}
\usepackage{hepnicenames}
\usepackage{enumerate}
\usepackage{soul}
\usepackage[normalem]{ulem}
\usepackage{comment}

\usepackage{units}

\usepackage{soul}

\makeatletter
\pretocmd{\chapter}{\addtocontents{toc}{\protect\addvspace{15\p@}}}{}{}
\pretocmd{\section}{\addtocontents{toc}{\protect\addvspace{5\p@}}}{}{}
\makeatother

\usepackage{mathrsfs,graphicx,rotating,amsmath,amsfonts,mathtools,booktabs,amssymb,wasysym}
\usepackage{hyperref}\usepackage{slashed}
\usepackage[nosort]{cite}
\usepackage[table,xcdraw,dvipsnames]{xcolor}
\usepackage{bm}
\usepackage{graphicx}
\usepackage{multirow,multicol}
\hypersetup{colorlinks,bookmarksopen,bookmarksnumbered,
	linkcolor=blus,pdfstartview=FitH,urlcolor=rossos,citecolor=verde}
\allowdisplaybreaks

\renewcommand{\[}{\left[}

\newcommand{\Lag}{\mathscr{L}}

\newcommand{\mio}[1]{}

\newcommand{\bpm}{\begin{pmatrix}}
\newcommand{\epm}{\end{pmatrix}}

\usepackage{mathrsfs}

\newcommand{\One}{1\!\!\hbox{I}}
\renewcommand{\One}{\mathbb{1}}

\renewcommand{\Im}{{\rm Im}\,}
\renewcommand{\Re}{{\rm Re}\,}
\allowdisplaybreaks
\usepackage{multicol}
\usepackage{color}
\definecolor{rosso}{cmyk}{0,1,1,0.4}
\definecolor{rossos}{cmyk}{0,1,1,0.55}
\definecolor{rossoc}{cmyk}{0,1,1,0.2}
\definecolor{blu}{cmyk}{1,1,0,0.3}
\definecolor{blus}{cmyk}{1,1,0,0.6}
\definecolor{bluc}{cmyk}{1,1,0,0.1}
\definecolor{verde}{cmyk}{0.92,0,0.59,0.25}
\definecolor{verdec}{cmyk}{0.92,0,0.59,0.15}
\definecolor{verdes}{cmyk}{0.92,0,0.59,0.4}

\newcommand{\eq}[1]{~{\rm (\ref{eq:#1})}}

\newcommand{\GeV}{\,{\rm GeV}}

\newcommand{\Tr}{\,{\rm Tr}}
\newcommand{\diag}{\,{\rm diag}}

\newcommand{\nn}{\nonumber}
\newcommand{\beq}{\begin{equation}}
\newcommand{\eeq}{\end{equation}}
\newcommand{\mb}[1]{\mbox{\boldmath $#1$}}
\newcommand{\bea}{\begin{eqnarray}}
\newcommand{\eea}{\end{eqnarray}}
\newcommand{\be}{\begin{equation}}
\newcommand{\ee}{\end{equation}}
\font\tenrsfs=rsfs10 at 12pt
\font\sevenrsfs=rsfs7
\font\fiversfs=rsfs5
\newfam\rsfsfam
\textfont\rsfsfam=\tenrsfs
\scriptfont\rsfsfam=\sevenrsfs
\scriptscriptfont\rsfsfam=\fiversfs

\def\nn{\nonumber}

\renewenvironment{thebibliography}[1]
{\begin{multicols}{2}[\section*{\refname}]%
		\@mkboth{\MakeUppercase\refname}{\MakeUppercase\refname}%
		\list{\@biblabel{\@arabic\c@enumiv}}%
		{\settowidth\labelwidth{\@biblabel{#1}}%
			\leftmargin\labelwidth
			\advance\leftmargin\labelsep
			\@openbib@code
			\usecounter{enumiv}%
			\let\p@enumiv\@empty
			\renewcommand\theenumiv{\@arabic\c@enumiv}}%
		\sloppy
		\clubpenalty4000
		\@clubpenalty \clubpenalty
		\widowpenalty4000%
		\sfcode`\.\@m}
	{\def\@noitemerr
		{\@latex@warning{Empty `thebibliography' environment}}%
		\endlist\end{multicols}}

\newcommand{\SU}{\,{\rm SU}}
\newcommand{\SL}{\,{\rm SL}}

\newcommand{\U}{\,{\rm U}}

\makeatletter
\font\ital=cmu10

\def\hhref#1{\href{http://arxiv.org/abs/#1}{arXiv:#1}}
\usepackage{xstring}
\newcommand{\hhrefq}[1]{\IfSubStr{#1}{:}{\href{http://inspirehep.net/search?ln=en&ln=en&p=#1&of=hb&action_search=Search&sf=&so=d&rm=&rg=25&sc=0}{InSpire:#1}}{\hhref{#1}}}

\def\art{\@ifnextchar[{\eart}{\oart}}
\def\eart[#1]#2#3#4#5#6{{\rm #2}, {\em #3 \bf #4} {\rm (#6) #5} ({\em #1})}
\def\article{\@ifnextchar[{\earticle}{\oarticle}}
\def\oarticle#1#2#3#4#5#6{{\rm #1}, {\ital `#6'}, {\rm #2 #3 (#5) #4}}
\def\earticle[#1]#2#3#4#5#6#7{{\rm #2}, {\ital `#7'}, {\rm #3 #4 (#6) #5}  [\hhrefq{#1}]}
\def\hepart[#1]#2{{\rm #2, \sl#1}}
\def\heparticle[#1]#2#3{#2, {\ital `#3'} [\hhrefq{#1}]}
\newcommand{\doi}[1]{\href{http://dx.doi.org/#1}{[link]}}

\newcommand{\hhrefqq}[1]{\IfBeginWith{#1}{10.}{\href{https://doi.org/#1}{doi:#1}}{\hhrefq{#1}}}

\renewenvironment{thebibliography}[1]
{\begin{multicols}{2}[\section*{\refname}]%
		\@mkboth{\MakeUppercase\refname}{\MakeUppercase\refname}%
		\list{\@biblabel{\@arabic\c@enumiv}}%
		{\settowidth\labelwidth{\@biblabel{#1}}%
			\leftmargin\labelwidth
			\advance\leftmargin\labelsep
			\@openbib@code
			\usecounter{enumiv}%
			\let\p@enumiv\@empty
			\renewcommand\theenumiv{\@arabic\c@enumiv}}%
		\sloppy
		\clubpenalty4000
		\@clubpenalty \clubpenalty
		\widowpenalty4000%
		\sfcode`\.\@m}
	{\renewcommand{\@noitemerr}
		{\@latex@warning{Empty `thebibliography' environment}}%
		\endlist\end{multicols}}

\newcounter{alphaequation}[equation]
\def\thealphaequation{\theequation\hbox to
	0.6em{\hfil\alph{alphaequation}\hfil}}
\def\eqnsystem#1{
	\def\@eqnnum{{\rm (\thealphaequation)}}
	\def\@@eqncr{\let\@tempa\relax \ifcase\@eqcnt \def\@tempa{& & &} \or
		\def\@tempa{& &}\or \def\@tempa{&}\fi\@tempa
		\if@eqnsw\@eqnnum\refstepcounter{alphaequation}\fi
		\global\@eqnswtrue\global\@eqcnt=0\cr}
	\refstepcounter{equation} \let\@currentlabel\theequation \def\@tempb{#1}
	\ifx\@tempb\empty\else\label{#1}\fi
	\refstepcounter{alphaequation}
	\let\@currentlabel\thealphaequation
	\global\@eqnswtrue\global\@eqcnt=0 \tabskip\@centering\let\\=\@eqncr
	$$\halign to \displaywidth\bgroup \@eqnsel\hskip\@centering
	$\displaystyle\tabskip\z@{##}$&\global\@eqcnt\@ne
	\hskip2\arraycolsep\hfil${##}$\hfil& \global\@eqcnt\tw@\hskip2\arraycolsep
	$\displaystyle\tabskip\z@{##}$\hfil
	\tabskip\@centering&\llap{##}\tabskip\z@\cr}
\def\endeqnsystem{\@@eqncr\egroup$$\global\@ignoretrue} \makeatother

\newcommand{\tS}{{\mathscr S}}
\newcommand{\tT}{{\mathscr T}}

\definecolor{Gray}{gray}{0.95}

\usepackage{tocloft}
\begin{document}
\thispagestyle{empty}
\vspace{0.1cm}
\begin{center}
{\LARGE \bf \color{rossos} Solving the strong CP problem without axions}\\[6ex]
{\bf\large Ferruccio Feruglio,$^a$ Matteo Parriciatu,$^{b,c}$ \\[1ex]
Alessandro Strumia,$^d$ Arsenii Titov$^d$}  \\[5mm]
{$^a$ \em INFN, Sezione di Padova, Italia}\\
{$^b$ \em INFN, Sezione di Roma Tre, Italia}\\
{$^c$ \em Dipartimento di Matematica e Fisica, Universit{\`a} di Roma Tre, Italia}\\
{$^d$ \em Dipartimento di Fisica, Universit{\`a} di Pisa, Italia}\\[4ex]
{\bf\large\color{blus} Abstract}
\begin{quote}
\large
We formulate general conditions under which the strong CP problem is solved by spontaneous CP violation. 
Quark-mass matrix elements are polynomials in the CP-breaking order parameters, 
engineered such that their determinant is a real constant. 
This scheme permits only a limited number of textures.
These conditions can be realized in supersymmetric theories with CP as an anomaly-free
local flavor symmetry, suggesting a unified solution to the strong CP problem and the flavor puzzle. 
Our solution can be implemented using either modular invariance or a local U(1) symmetry.
We present modular-invariant realizations where matter fields are assigned small modular weights $\pm2$ ($\pm1$),
utilising higher levels $N=2$ ($N=3$). 
Heavy quarks are in general not required, but their
presence allows for models where colored particles fill non-singlet
representations of the flavor group.
\end{quote}
\end{center}

\clearpage
\thispagestyle{empty}
{\setstretch{1.2}\tableofcontents}
\newpage

\section{Introduction}
The strong CP puzzle  in quantum chromodynamics (QCD) consists in explaining the measured smallness $ | \bar\theta|\lesssim 10^{-10}$~\cite{2001.11966}
of the re-phasing invariant combination 
\beq\bar\theta= \theta_{\rm QCD} + \arg\det m_q\eeq
of different terms in the QCD Lagrangian: the $\theta_{\rm QCD}$ angle and the quark mass matrix $m_q$,
\beq
\Lag_{\rm QCD} = \bar q (i \slashed{D} - m_q)q - \frac{\Tr \, G^2}{4g_{\rm s}^2}  +\frac{ \theta_{\rm QCD} }{32\pi^2} \Tr\, G\tilde{G}.\eeq
The puzzle arises because $m_q$ must also reproduce the CKM phase, observed to be large, $\delta_{\rm CKM}\sim 1$. 
So one naively expects a large $\arg\det m_q$.

  To address the strong CP problem, one approach involves transforming $\bar{\theta}$ from a fixed parameter into a light, dynamical field known as the axion. This field dynamically adjusts itself to cancel any potentially large contributions from $\arg\det m_q$. This mechanism can be realized by assuming a spontaneously broken global U(1) symmetry with QCD anomalies, resulting in the axion emerging as a pseudo-Goldstone boson. The symmetry breaking scale must be significantly higher than the weak scale to prevent undetected effects. Numerous experiments are striving to achieve the sensitivity required to detect the axion~\cite{2003.01100}. Additionally, the axion is a promising candidate for dark matter.

\smallskip

Other possible solutions to the strong CP problem do not involve the axion nor any new physics at low energy.
The idea is enforcing P or CP invariance on the fundamental theory,
which implies $\bar\theta=0$.
The challenge lies in spontaneously breaking P or CP in a way that maintains a small $\arg\det m_q$ 
while allowing for a large $\delta_{\rm CKM}$. 
In the Standard Model (SM), both P and CP are explicitly broken, 
so preserving invariance under these discrete symmetries requires an extension of the SM.
\begin{itemize}
\item A {\bf P-invariant} theory requires a drastic extension of the SM.
Parity can be enforced by embedding the SM gauge group into a left-right extension, and/or doubling the fermion content of the SM
by introducing mirror fermions. 
 In scenarios without mirror fermions, parity requires that quark Yukawa couplings are described by Hermitian matrices. 
 A set of Higgs fields is then needed to break the extended gauge group. 
 If the vacuum expectation values (VEVs) of these Higgs fields are real, 
 the resulting quark mass matrices are Hermitian, leading to $\arg\det m_q = 0$ and an unconstrained CKM phase at tree level~\cite{Beg:1978mt,Mohapatra:1978fy}. 
However, in the presence of explicit CP violation, the Higgs VEVs
are expected to be complex and the quark mass matrices are no longer hermitian.
Additional constraints, such as supersymmetry, have to be added to avoid this scenario \cite{Mohapatra:1995xd}.
Alternatively, when mirror fermions are present, 
a cancellation between the contribution
of ordinary and mirror fermions can ensure $\arg\det m_q=0$ at  tree level before P breaking \cite{Barr:1991qx,Craig:2020bnv,Bonnefoy:2023afx}.

\item A {\bf CP-invariant} theory does not require a drastic extension of the SM, as
in the SM CP is broken only by the Yukawa interactions. If Yukawa couplings are promoted
to dynamical variables, a possibility favored by string theory, CP violation can arise spontaneously as a dynamical effect. 
So far this scenario has mainly been realized through the mixing 
of quarks with ultra-heavy vector-like quarks via complex couplings. 
According to the original proposal by Nelson and Barr~\cite{Nelson:1983zb,Barr:1984qx},
this mixing generates a CKM phase at tree level, while a judicious choice of Yukawa couplings, possibly enforced by symmetry requirements, 
forbids any tree-level contribution to $\arg\det m_q$.
In a supersymmetric realization, as long as supersymmetry remains exact, non-renormalization theorems
protect this result from quantum corrections~\cite{Ellis:1982tk}. 
A non-vanishing $\delta_{\rm CKM}$ can arise both from the coupling
between ordinary and heavy quarks and from a loop-induced wave function renormalization of the quark fields~\cite{Hiller:2001qg,Hiller:2002um}. 
For further works on the solutions to the strong CP problem 
based on spontaneous CP violation, see e.g.~\cite{1307.0710,1412.3805,1506.05433,2105.09122,2106.09108,2406.01260}.

\end{itemize}
A recent paper~\cite{Feruglio:2023uof} demonstrated that the observed CKM phase together with
$\arg\det m_q=0$ can be reproduced
in a minimal setting, where no extra quarks are introduced and CP 
is spontaneously broken by the VEV of a single complex field, the modulus. 
This structure  naturally arises within CP and modular-invariant supersymmetric theories. 
In such theories, fermions undergo a local chiral rotation under the action of the modular group.
The absence of mixed chiral/$\SU(3)_c$ anomalies 
requires a specific choice of the modular transformation laws of matter fields, which ensures $\arg\det m_q=0$ at tree level.
CP and modular invariance are key ingredients in string theory compactifications~\cite{Hamidi:1986vh,Dixon:1986qv,Lauer:1989ax,Lauer:1990tm} and 
are generally inherited by the low-energy supergravity theory, 
where this solution to the strong CP problem can be  implemented. 

\smallskip

The aim of the present paper is to identify general conditions under which spontaneous CP violation can solve the strong CP problem. 
We also look for the most general theoretical framework  where such conditions are realized. 
Starting from a toy model in section~\ref{toytwogen}, 
and considering a realistic set-up in section~\ref{threegen},
we formulate these conditions in section~\ref{gencon}. 
In section~\ref{sec:Yukpatterns} we classify all possible
patterns of Yukawa couplings that can be realized in the absence of extra heavy quarks.

The conditions we are investigating naturally arise in a supersymmetric theory where the SM gauge group 
is extended to include a local flavor symmetry, as discussed in section \ref{thsetup}. 
Two examples of such flavor symmetry --- $\U(1)$ and modular invariance --- are considered in section~\ref{subex}.
The conditions of absence of mixed flavor/SM anomalies,
examined in section \ref{anomalies}, single out a set of flavor quantum numbers that are instrumental in getting
$\arg\det m_q=0$. 

When heavy vector-like quarks are present, our general conditions can be implemented in a variety of different ways.
These are illustrated in section~\ref{HQ}, where we discuss the corresponding low-energy effective theory and the mechanism of gauge anomaly cancellation.
In particular, we show that the Nelson-Barr class of models is a particular case of a more general framework. 

\medskip

Finally, in sections~\ref{cmod}, \ref{N=2}, \ref{N=3}, we
exhibit new models based on modular invariance.
In~\cite{Feruglio:2023uof} the simplest model utilized singlet representations of the full modular group $\SL(2,\mathbb{Z})$,
with modular weights $\{-6, 0, 6\}$ for the three generations.
Here, we consider for the first time models with lowest weights $\pm 1$
($\pm2$) based on singlet representations of
the infinite principal congruence 
sub-group $\Gamma(3)$ ($\Gamma(2)$) of $\SL(2,\mathbb{Z})$.

Furthermore, the flavor symmetry can be embedded in a non-trivial way within the modular group, 
such that quarks 
fill higher-dimensional representations of non-abelian finite quotient groups such as 
$\Gamma_N \equiv \SL(2,\mathbb{Z})/\Gamma(N)$
with $N \geq 2$.
Recently~\cite{2404.08032} (see also~\cite{2404.00858}) found that such models cannot solve the strong CP problem 
if based on light quarks only.
This limitation arises because it is not possible to reproduce the observed quark masses and mixings given the small number of free parameters.
Only the models utilising one-dimensional $\Gamma_N$ representations were found to be viable, and
these models also involve high weights of the matter fields. 
Here we show that, by introducing heavy quarks, it is possible to construct non-abelian realizations with low weights,
that solve the strong CP problem while reproducing quark masses and mixings.
Section~\ref{N=2} presents models based on $\Gamma_2$ with weights $\pm2$, and section~\ref{N=3} 
explores models based on $\Gamma_3$ with weights $\pm1$.
While heavy quarks are not a necessary component of our solutions to the strong CP problem, 
they can effectively mimic the towers of heavy degrees of freedom that are present in string theory compactifications.

\section{The general idea}
We consider a theory invariant under CP, where
CP is spontaneously broken by a set of generic  spin-zero gauge-invariant local operators $z$, which can be either elementary or composite scalars.
Without loss of generality, we can assume that $z$ are dimensionless.
It is not difficult to provide examples of mass matrices $m_q$, within the quark content of the SM, such that $\arg\det m_q=0$. 
We start by providing simple toy examples before moving to the general structure.

\subsection{A toy model with two generations \label{toytwogen}}
To this purpose, it is useful to start from a toy model with two fermion generations
and a single operator $z$. 
The Yukawa couplings in the quark sector involve electroweak singlet quarks $D^c_i,U^c_i$, 
the weak doublets $Q_i$ and the Higgs doublet $H$ (two Higgs doublets $H_{u,d}$ in supersymmetric theories)  acquiring a real VEV. 
We assume that the quark mass matrices are of the type:
\be\label{toy}
m_d=
\left(
\begin{array}{cc}
d_{11} z^{k^d_{11}}&d_{12}z^{k^d_{12}}\\
d_{21} z^{k^d_{21}}&d_{22}z^{k^d_{22}}\\
\end{array}
\right),
\qquad
m_u=
\left(
\begin{array}{cc}
u_{11} z^{k^u_{11}}&u_{12}z^{k^u_{12}}\\
u_{21} z^{k^u_{21}}&u_{22}z^{k^u_{22}}\\
\end{array}
\right),
\ee
where $d_{ij}$ and $u_{ij}$ are constants, required to be real by CP invariance.
The determinant of the full quark mass matrix $m=m_d \oplus m_u$ is
\be\nn
\det m=\left(d_{11} d_{22} z^{k^d_{11}+k^d_{22}}-d_{12} d_{21} z^{k^d_{12}+k^d_{21}}\right)\left(u_{11} u_{22} z^{k^u_{11}+k^u_{22}}-u_{12} u_{21} z^{k^u_{12}+k^u_{21}}\right).
\ee
We look for the conditions under which $\arg\det m=0$. This is realized if $\det m$ is a positive 
non-vanishing constant. More generally, the requirement that $\det m$ is proportional to a power of $z$
for generic parameters $d_{ij}$ and $u_{ij}$ restricts the powers as 
\begin{align}\label{c1}
k_{11}^d+ k_{22}^d = k^d_{12}+k^d_{21},\qquad
k^u_{11}+ k^u_{22} = k^u_{12}+k^u_{21}.
\end{align}
An interesting class of solutions to the above equations is\footnote{This is not the most general solution. For instance, it cannot cover the case $k^d_{ij}=k^u_{12}=k^u_{21}=0$, $k^u_{11}=-k^u_{22}\ne 0$.}
\beq
k_{ij}^d=k_{D^c_i}+k_{Q_j}+k_{H_d},\qquad
k^u_{ij}=k_{U^c_i}+k_{Q_j}+k_{H_u}.
\eeq
Notice that the $k$ are determined up to additive constants, as those of Higgs multiplets can be shifted
as $k_{H_q} \to k_{H_q}+\Delta_{H_q}$ and those of left-handed quarks as $k_{Q_i}\to k_{Q_i}+\Delta_{Q}$ 
provided that those of right-handed quarks are simultaneously shifted as
\beq \label{shift1}k_{D^c_i}\to k_{D^c_i} - \Delta_{H_d} - \Delta_{Q},\qquad
k_{U^c_i}\to k_{U^c_i} - \Delta_{H_u} - \Delta_{Q}.\eeq
The parameters $k_{D^c_i},k_{U^c_i},k_{Q_j},k_{H_d},k_{H_u}$ will later be interpreted as charges in models with U(1) symmetries, 
and as weights in models with modular symmetries.
This simple model can be justified, for instance, by enforcing a U(1) family symmetry, 
spontaneously broken by a complex scalar field $z$ carrying a positive unit $k_z=+1$ of the charge in a supersymmetric realization 
where $\Phi=\{D^c,U^c,Q,H_{u,d}\}$ are superfields.
We use a different sign conventions of matter and $z$ charge:
\beq\nn
\Phi_i\to e^{ -i k_{\Phi_i}\alpha }\Phi_i,
\qquad 
z\to e^{ + i k_z\alpha}z.
\eeq
 With the above parametrisation the determinant of the quark mass matrix reads
\beq\nn
\det m=\left(d_{11} d_{22} -d_{12} d_{21} \right)\left(u_{11} u_{22} -u_{12} u_{21} \right)
z^d,
\eeq
where
\beq\label{sum}
d=\sum_{i=1}^2(k_{D^c_i}+k_{U^c_i}+2 k_{Q_i}+ k_{H_d}+k_{H_u})=\sum_{i=1}^2(k^d_{ii}+k^u_{ii})\,.
\eeq
A $z$-independent real determinant is obtained if $d=0$.

\medskip

When the constraint $d=0$ is realized by a non-trivial choice of $k_{D^c_i},k_{U^c_i},k_{Q_j},k_{H_d},k_{H_u}$, 
some exponents $k^d_{ij}$ or $k^u_{ij}$ are necessarily negative.\footnote{Here we dismiss the trivial case where all exponents vanish. We assume that all exponents can vanish either in the down or in the up quark sector, but not in both.}
For example, assuming $k_{H_{u,d}}=0$, $k_{D^c_1}=k_{U^c_1}=-k_{Q_1}>0$, $k_{D^c_2}=k_{U^c_2}=-k_{Q_2}<0$
we have
\be\nn
m_d=
\left(
\begin{array}{cc}
d_{11} &d_{12}~ z^k\\
d_{21}~ z^{-k}&d_{22}\\
\end{array}
\right),
\qquad
m_u=
\left(
\begin{array}{cc}
u_{11} &u_{12}~ z^k\\
u_{21}~ z^{-k} &u_{22}\\
\end{array}
\right)
\ee
where $k=k_{D^c_1}+k_{Q_2}>0$.
The $(m_q)_{21}$ entries are singular in the limit $z=0$. 
In an effective field theory (EFT), we expect singularities to occur when some state that has been eliminated from the
low-energy description becomes accidentally massless. 
An example in this sense will be presented in section~\ref{ex1}. 
If such an extra state does not exist in the full theory, the coefficients $d_{21}$ and $u_{21}$ must vanish.

\subsection{A model with three generations and the CKM phase}
\label{threegen}
In this toy model, we get $\arg\det m=0$ through the condition $d=0$ 
and an extra assumption about the sign
of $\det m$.  
The  model of section \ref{toytwogen} does not exhibit any physical phase, needed to describe $\delta_{\rm CKM}$, because no observable
CP-violating phase survives when there are only two generations. 

However, there is no CKM-like CP-violating phase also in the generic case of $N>2$ generations, 
when the mass matrices $m_u$ and $m_d$ are of the type\footnote{Without losing generality we set
$k_{H_{u,d}}=0$ by exploiting the freedom of eq.~(\ref{shift1}).}
\be
(m_d)_{ij}=d_{ij}~ z^{k_{D^c_i}+k_{Q_j}},
\qquad
(m_u)_{ij}=u_{ij}~ z^{k_{U^c_i}+k_{Q_j}},
\ee
with real constants $d_{ij}$ and $u_{ij}$ and a single operator $z$, 
see e.g.~\cite{0704.0697}.
Indeed, the mass matrices depend on a single phase, $\varphi=\arg z$, that can be removed through the field redefinition:\footnote{We assume these transformations are not anomalous, 
see section \ref{anomalies}. If the transformations are local, the VEV of $z$ can be made real by a gauge choice.}
\be
\label{fred}
D^c_i\to e^{-i k_{D^c_i} \varphi} D^c_i,
\qquad
U^c_i\to e^{-i k_{U^c_i} \varphi} U^c_i,
\qquad
Q_i\to e^{-i k_{Q_i} \varphi}  Q_i.
\ee
We can try to avoid this negative conclusion by turning on a dependence of the quark mass matrices on the conjugate 
field $\bar z$. In the framework of a spontaneously broken U(1) flavor symmetry, 
where $z$ carries a charge $k_z=+1$, it would be natural to assign charge $-1$ to $\bar z$.
If we allow the entries of the quark mass matrices to depend on the powers of the conjugate 
field $\bar z$, all entries with a negative charge $k_{D^c_i}+k_{Q_j}$,
could be filled by terms proportional to $\bar z^{-(k_{D^c_i}+k_{Q_j})}$. 
However, with a single complex field $z$, even this extension contains no physical phases, 
since the field redefinition of eq.~(\ref{fred}) keeps removing all phases.
Phases coming from the transformation needed to put kinetic terms into the canonical forms are not expected to
change this conclusion. 
A non-diagonal Kahler kinetic term of the type $Q^\dagger_i Q_j$ has an overall charge $k_{Q_j}-k_{Q_i}$,
which can be neutralized by a factor $z^{k_{Q_j}-k_{Q_i}}$ when $k_{Q_j}-k_{Q_i}>0$ or
by $\bar z^{k_{Q_i}-k_{Q_j}}$ when $k_{Q_j}-k_{Q_i}<0$. 
In any case the transformation in eq.~(\ref{fred}) eliminates the dependence on the phase of $z$. These conclusions are independent of the number of generations.
The general reason is that one $z$ breaking a U(1) does not lead to any physical phase: U(1)-invariance allows to rotate $z$ to real values.
 
\medskip

If more than one complex operators $z_1,z_2,\ldots $ are present, different generalisations of the above model can be formulated. 
For example, following the analogy with the flavor symmetry realization, the flavor symmetry can be extended from
U(1) to $\U(1)_1\otimes\U(1)_2$, broken by two scalar fields $z_1$ and $z_2$ carrying charges $(1,0)$ and $(0,1)$.
Now the quark mass matrices are
\be
(m_d)_{ij}=d_{ij}~ z_1^{k^1_{D^c_i} +k^1_{Q_j}}~z_2^{k^2_{D^c_i} +k^2_{Q_j}},
\qquad
(m_u)_{ij}=u_{ij}~ z_1^{k^1_{U^c_i}+k^1_{Q_j}}~z_2^{k^2_{U^c_i}+k^2_{Q_j}}.
\ee 
However, also in this case no physical phase survives in the mass matrices since phases can be eliminated by the transformation:
\begin{align}
\label{fred2}
D^c_i\to &~e^{-i k^1_{D^c_i} \varphi_1-i k^2_{D^c_i}  \varphi_2}~ D^c_i\nn\\
U^c_i\to &~e^{-i k^1_{U^c_i} \varphi_1-i k^2_{U^c_i} \varphi_2}~ U^c_i\\
Q_i\to &~e^{-i k^1_{Q_i} \varphi_1-i k^2_{Q_i} \varphi_2}~ Q_i.\nn
\end{align}
The general reason is that each $\U(1)_i$ allows to rotate to real values its associated $z_i$.

So, a successful generalization consists in assuming a single U(1) flavor symmetry broken by more than one operator $z$.
In such a case the relative phases among the different $z$ are physical.
Consider for example a model with two operators $z_1$ and $z_2$ carrying U(1) charges $+1$ and $+2$, respectively. 
When $k_{D^c_i}+k_{Q_j}$ is positive, the entry $ij$ of the mass matrix $m_d$ is a polynomial in $z_1$ and $z_2$.
For instance, if $k_{D^c_1}+k_{Q_1}=3$ we have $(m_d)_{11}=a_{11}~ z_1^3+a_{11}'~ z_1 z_2$. 
Since the matrix elements are no longer simple monomials in the variables $z_1$ and $z_2$, 
we do not expect that all phases can be eliminated by field redefinitions.
The relative phase in $z_1^2/z_2$ is now physical.
We can postulate that for negative $k_{D^c_i}+k_{Q_j}$ the entry $ij$ of the mass matrix vanishes, unless 
the entry is generated
by integrating out states that become singular when either $z_1$ or $z_2$ or both go to zero, as will be shown in section~\ref{ex1}.
It is not difficult to find the general conditions under which the determinant of the quark mass matrix in a similar construction
is real, which we discuss in the next section.

\subsection{The general solution to the strong CP problem}\label{gencon}
We identify a general framework where the determinant of the quark mass matrices is real and a non-trivial CKM phase is present. 
We assume invariance under CP and under the $\SU(3)_c\otimes\SU(2)_L\otimes\U(1)_Y$ gauge group,
spontaneously broken by one or two Higgs doublets $H_{u,d}$, giving rise to Yukawa couplings as in the SM or in the MSSM, respectively. 
If two doublets are present, we assume their VEVs are real.\footnote{For a discussion within the MSSM, see~\cite{Maekawa:1992un,Pomarol:1992bm,Branco:1993js}.}
Thus, when analyzing the quark mass matrices and their phases, we can write the Yukawa Lagrangian as
\be
\Lag_{\rm Yuk}=U^c_i  Y^u_{ij} Q_j~H_u + D^c_i Y^d_{ij}~ Q_j~H_d + \cdots
\ee
where $\cdots$ stand for the leptonic contributions.
We are led to assume:
\begin{itemize}
\item[1.]  At tree level, the Yukawa couplings $Y^q_{ij}$ $(q=u,d)$ of up and down quark sectors are polynomials in the complex variables
$z_a$  $(a=1,...,N)$ with real coefficients. The $z_a$ are VEVs of dimension-less spin-zero local operators, which can be either elementary or composite. 
They are responsible for the spontaneous breaking  of CP.
The assumption that $Y^q_{ij}$ is a polynomial ---  the sum of $z_1^{\alpha_1}...~z_N^{\alpha_N}$ monomials 
with non-negative exponents $\alpha_a$ --- 
can be justified in an EFT framework assuming that the full theory contains no states that become massless, generating singularities. 
The assumption of real coefficients can be justified by imposing CP invariance broken by $z_a$.

\item[2.] Each variable $z_a$ is characterized by a positive weight $k_a$ and the polynomials $Y^q_{ij}(z)$ are weighted-homogeneous of degree $d^q_{ij}$: 
for any complex $\lambda$
\begin{align}
\label{quasihom}
Y^q_{ij}(\lambda^{k_1}z_1, \ldots ,\lambda^{k_N} z_N)=\lambda^{d^q_{ij}}Y^q_{ij}(z_1, \ldots ,z_N)~~~,
\end{align}
where $d^q_{ij}$ are real numbers.
\item[3.] The determinant of $Y^q_{ij}$ is also a weighed-homogeneous polynomial  in the variables $z_a$ of degree $d_q\equiv \sum_i d^q_{ii}$
\begin{align}
\det \left[Y^q_{ij}(\lambda^{k_1}z_1,...,\lambda^{k_N} z_N)\right]=\lambda^{d_q} \det\left[Y^q_{ij}(z_1,...,z_N)\right].
\end{align}

\item[4.] The total degree vanishes: 
\be
d=d_u+d_d=0.
\ee
\end{itemize}
\vskip 0.3 cm
A first consequence of these assumptions is that $Y^q_{ij}=0$ whenever $d^q_{ij}<0$. Indeed, from point 1.~it follows that any matrix element in $Y_{ij}$ is a combination of monomials of the type
\be\nn
z_1^{\alpha_1} \cdots z_N^{\alpha_N}~,
\ee
with non-negative exponents $\alpha_a$, and we have 
\be\nn
(\lambda^{k_1}z_1)^{\alpha_1} \cdots (\lambda^{k_N}z_N)^{\alpha_N}=\lambda^{d^q_{ij}}z_1^{\alpha_1}\cdots z_N^{\alpha_N}\qquad\hbox{where}\qquad
d^q_{ij}= k_1\alpha_1+ \cdots +k_N\alpha_N\ge 0
\ee
is non-negative by assumption.
Therefore, for those entries $Y^q_{ij}$, such that $d^q_{ij}<0$, the only way to satisfy eq.~(\ref{quasihom}) is $Y^q_{ij}=0$. 
As in our toy model, non-vanishing entries $Y^q_{ij}$ with $d^q_{ij}<0$ 
can be generated in a low-energy effective theory
by integrating out extra heavy quarks that become massless when some variable $z_a$ goes to zero.
Likewise, when the quark content coincides with that of the SM, the determinant of $Y^q_{ij}$ either has a non-negative degree $d_q$ or vanishes, both in the up and down sectors. 
It follows that the condition  4.\ can only be satisfied by having simultaneously 
\be\label{simul}
d_d=d_u=0.
\ee
This result can be avoided if extra quarks are present and the low-energy effective theory, obtained by integrating out
the extra degrees of freedom, exhibits singularities when some $z_a$ goes to zero. 
In this case condition 4.~can be realized with $d_d$ and $d_u$ of different signs.

\medskip

The condition 3. is automatically satisfied if
\be
\label{sol1}
d^d_{ij}=k_{D^c_i}+k_{Q_j}+k_{H_d},\qquad
d^u_{ij}=k_{U^c_i}+k_{Q_j}+k_{H_u},
\ee
as we can see from a direct check: 
\begin{align}
&{\det}\left[Y^u_{ij}(\lambda^{k_1}z_1, \ldots ,\lambda^{k_N} z_N)\right]=
{\det}\left[\lambda^{d^u_{ij}}Y^u_{ij}(z_1,\ldots ,z_N)\right]=
\lambda^{d_u}{\det}\left[Y^u_{ij}(z_1,\ldots,z_N)\right],
\end{align}
where
\be
d_u=\sum_{i=1}^3(k_{U^c_i}+k_{Q_i})+3 k_{H_u}.
\ee
A similar relation holds in the down-quark sector.
A consequence of 2., 3.\ and 4.\ is that the determinant of $Y^{u,d}$ is a real constant. 
The requirement 2. implies the relation
\be
\label{pde}
k_a z_a \frac{\partial Y_{ij}(z_1, \ldots ,z_N)}{\partial z_a}=d_{ij} Y_{ij}(z_1, \ldots ,z_N)~.
\ee
This is derived starting from the identity
\be
k_a z_a \frac{\partial Y_{ij}(\lambda^{k_1}z_1, \ldots,\lambda^{k_N} z_N)}{\partial z_a}=\lambda\frac{\partial Y_{ij}(\lambda^{k_1}z_1,\ldots ,\lambda^{k_N} z_N)}{\partial\lambda}
\nn
\ee 
and using eq. (\ref{quasihom}) we get
 \be
 k_a z_a \frac{\partial Y_{ij}(\lambda^{k_1}z_1,\ldots,\lambda^{k_N} z_N)}{\partial z_a}=\lambda\frac{\partial \lambda^{d_{ij}}Y_{ij}(z_1, \ldots,z_N)}{\partial\lambda}=\lambda~d_{ij}~\lambda^{d_{ij}-1}Y_{ij}(z_1,\ldots,z_N).
 \nn
 \ee
Setting $\lambda=1$  gives eq.~(\ref{pde}).
From the requirements 3.\ and 4.\ (realized as in eq.~(\ref{simul})) it follows that the determinant of $Y^q$ $(q=d,u)$ is a polynomial of vanishing degree:
\begin{align}
{\det}\left[Y^q_{ik}(\lambda^{k_1}z_1,\ldots ,\lambda^{k_N} z_N)\right]=
\lambda^{d_q}~{\det}\left[Y^q_{ik}(z_1, \ldots,z_N)\right]={\det}\left[Y^q_{ik}(z_1, \ldots,z_N)\right]\nn\,,
\end{align}
which implies
\begin{align}
\label{w0}
k_a \frac{\partial \det Y^q_{ij}}{\partial \ln z_a}=0.
\end{align}
The determinant is a sum of monomials of the type $z_1^{\alpha_1}\cdots z_N^{\alpha_N}$, with
non-negative $\alpha_1, \ldots,\alpha_N$. We have
\be
k_a  \frac{\partial z_1^{\alpha_1}\cdots z_N^{\alpha_N}}{\partial \ln z_a}=k_a \alpha_a~z_1^{\alpha_1}\cdots z_N^{\alpha_N}.
\ee
Barring accidental cancellations among the free parameters of $Y^{d,u}$, each term should independently vanish, implying $k_a \alpha_a=0$. From the positivity of the weights, it follows that the only solution is $\alpha_a=0$, so the determinant is a constant.

Since all its coefficients are real, it is a real constant.
For this to work, it is essential that all weights $k_a$ are positive. If a variable $z_0$ had vanishing weight, then
a determinant depending only on $z_0$ would satisfy eq.~(\ref{w0}). 
If negative weights are allowed, we can easily
build examples of non-constant determinants. Consider two variables $z_1,z_2$ with weights $+1$ and $-2$, respectively. Then $z_1^2 z_2$
would be a polynomial with a vanishing degree. Notice that the dependence of $Y^q_{ij}$ on the conjugate variables $\bar z_a$
is forbidden, if $\bar z_a$ are assigned negative weights. Nevertheless, in section \ref{NB} we will see how a dependence on 
$\bar z_a$ of the quark mass matrices can arise from the functional integration over a heavy sector of the theory.

\subsection{Patterns of Yukawa matrices for $\bar\theta=0$, $\delta_{\rm CKM}\neq 0$}\label{sec:Yukpatterns}
When all above conditions are fulfilled, only a limited number of independent patterns of Yukawa couplings  can be realized. 
Here we classify the non-singular patterns arising when no extra quarks are included.
To allow for three massive up and down quark generations, we require in each sector
\be
\label{det}
\det[Y_{ij}(z_1,\ldots,z_N)]=\hbox{constant}\ne 0.
\ee
The structure of $Y_{ij}(z_1,\ldots,z_N)$ is
\beq
\left(
\begin{array}{ccc}
Y_{11}^{(p_1+q_1)}&Y_{12}^{(p_1+q_2)}&Y_{13}^{(p_1+q_3)}\\
Y_{21}^{(p_2+q_1)}&Y_{22}^{(p_2+q_2)}&Y_{23}^{(p_2+q_3)}\\
Y_{31}^{(p_3+q_1)}&Y_{32}^{(p_3+q_2)}&Y_{33}^{(p_3+q_3)}\\
\end{array}
\right),
\eeq
where for each entry we have displayed the degree of the weighted-homogeneous polynomial. To reproduce eq.~(\ref{det}),
we need at least three entries $Y_{ij}$ constant and non-vanishing. By reordering rows and columns of $Y_{ij}$, without losing generality, we can assume that these entries are
$Y_{11}$, $Y_{22}$, and $Y_{33}$. This implies $p_i=-q_i$ $(i=1,2,3)$. We end up with:
\beq
\left(
\begin{array}{ccc}
Y_{11}^{(0)}&Y_{12}^{(p_1-p_2)}&Y_{13}^{(p_1-p_3)}\\
Y_{21}^{(p_2-p_1)}&Y_{22}^{(0)}&Y_{23}^{(p_2-p_3)}\\
Y_{31}^{(p_3-p_1)}&Y_{32}^{(p_3-p_2)}&Y_{33}^{(0)}\\
\end{array}
\right).
\eeq
We are left with two independent degrees, which we can choose as $p_1-p_2$ and $p_2-p_3$.
\begin{itemize}
\item[$\bullet$]
If they are both vanishing, all degrees also vanish and the entries $Y_{ij}$ are all non-vanishing constants.
This case is realized when $p_1=p_2=p_3$. 
Due to the equivalence relation of eq.~(\ref{shift1}), it is not restrictive to assume these weights are all vanishing. We get:
\be\label{caso1}
\left(
\begin{array}{ccc}
Y_{11}^{(0)}&Y_{12}^{(0)}&Y_{13}^{(0)}\\
Y_{21}^{(0)}&Y_{22}^{(0)}&Y_{23}^{(0)}\\
Y_{31}^{(0)}&Y_{32}^{(0)}&Y_{33}^{(0)}\\
\end{array}
\right).
\ee
\item[$\bullet$]
If one of $p_1-p_2$ and $p_2-p_3$ vanishes, we have a pattern with 5 non-vanishing constant entries, 2 vanishing entries and
2 non-trivial polynomials. For example, if $p_2=p_1\ne p_3$ we get
\be\label{caso2}
\left(
\begin{array}{ccc}
Y_{11}^{(0)}&Y_{12}^{(0)}&Y_{13}^{(p_1-p_3)}\\
Y_{21}^{(0)}&Y_{22}^{(0)}&Y_{23}^{(p_1-p_3)}\\
0&0&Y_{33}^{(0)}\\
\end{array}
\right).
\ee
In this specific case we have considered $p_1=p_2>p_3$. If $p_1=p_2<p_3$, the resulting pattern is the transposed
of the one in eq.~(\ref{caso2}). 
By permuting rows and columns it is always possible to match one of these two textures.

\item[$\bullet$]
If both $p_1-p_2$ and $p_2-p_3$ are non-vanishing (and their sum does not vanish, otherwise we go back to the previous case) we end up with
a pattern with 3 non-vanishing constant entries, 3 vanishing entries and
3 non-trivial polynomials, such as for example
\be\label{caso3}
\left(
\begin{array}{ccc}
Y_{11}^{(0)}&Y_{13}^{(p_1-p_2)}&Y_{13}^{(p_1-p_3)}\\
0&Y_{22}^{(0)}&Y_{23}^{(p_2-p_3)}\\
0&0&Y_{33}^{(0)}\\
\end{array}
\right).
\ee
It is not restrictive to label the weights $p_i$ such that $p_1>p_2>p_3$. Thus, up to permutations of rows and columns,
eq.~(\ref{caso3}) is the most general pattern when there are three distinct weights.
Due to the equivalence relation of eq.~(\ref{shift1}),
it is not restrictive to assume, for instance, $p_2=0$. Therefore we have three weights $p_i$, one vanishing,
one positive and one negative. 
This kind of texture has been considered in~\cite{Barr:1996wx}.
\item[$\bullet$]
Without losing generality we can assume that both $Y^d$ and $Y^u$ have the same pattern. Indeed, if in one sector we have
charges $k_{D^c_i}=-k_{Q_i}$, to get a non-vanishing constant determinant in the other sector we should similarly have $k_{U^c_\alpha}=-k_{Q_1}$,  $k_{U^c_\beta}=-k_{Q_2}$ and $k_{U^c_\gamma}=-k_{Q_3}$, with $\alpha\ne \beta\ne \gamma$. It suffices to reorder the rows 
to arrange $k_{U^c_i}=k_{D^c_i}=-k_{Q_i}$, thus obtaining the same texture in the two sectors.
\end{itemize}
In summary, when the theory fulfills the requirements 1., 2., 3., 4.\ and no quark mass vanishes, the matrices of Yukawa 
couplings of the up and down quark sectors have the same pattern, which coincides with one of the three cases shown in 
eq.s (\ref{caso1}-\ref{caso3}). 
CP is spontaneously broken by the VEVs of the complex scalar operators $z_a$, 
the topological angle $\theta_\mathrm{QCD}$ vanishes, and the determinant of the quark mass matrix is real. 
If this determinant is positive we end up with $\bar\theta=0$, at the tree level.
On the other hand, a non-vanishing CKM phase can originate in several ways. Either from the phases of the
$z_a$ VEVs, if the mass matrices have the pattern in eq.~(\ref{caso2}) or (\ref{caso3}), or from the phases occurring
in the special unitary transformations needed to put kinetic terms in canonical form.

\section{QFT realization with supersymmetry}\label{thsetup}
In this section we define a general setup where the assumptions outlined in the previous section are
satisfied. The assumption that $Y^q_{ij}$ is a polynomial in $z_a$, but not in the conjugate variables $\bar z_a$ calls for a supersymmetric realization. 
Since supersymmetry is not an exact symmetry of nature, supersymmetry-breaking effects should be accounted for. 
Real coefficients of the polynomial $Y^q_{ij}$ can be guaranteed by the CP invariance of the theory.
Focussing on ${\cal N}=1$ rigid supersymmetry, we impose invariance under CP and $G_{\rm SM}=\SU(3)_c\otimes\SU(2)_L\otimes\U(1)_Y$ gauge transformations, 
with a field content including the $G_{\rm SM}$ vector super-multiplets;
three generations of chiral multiplets $Q_i$, $U^c_i$, $D^c_i$, $L_i$, $E^c_i$ describing quarks and leptons; 
chiral multiplets $H_u$, $H_d$ describing two electroweak Higgs doublets with opposite hypercharge; 
a set of dimensionless, gauge-invariant chiral multiplets $Z=(Z_1,...,Z_N)$ which can be either
elementary or composite. 
In a standard superfield notation, the relevant part of the Lagrangian, including the strong gauge interactions, reads
\be\label{susylag}
\Lag=\int d^2\theta \, d^2\bar\theta~ K(e^{2V} \Phi,\Phi^\dagger) + \left[\int d^2\theta~ W(\Phi) +\hbox{h.c.}\right] +
\left[\frac{1}{16}\int d^2\theta~ f(\Phi)  W^a W^a+\hbox{h.c.}\right],
\ee 
where  $M_i=\{U^c_i,D^c_i, E^c_i, Q_i, L_i, H_{u,d}\}$ collectively denotes the matter and Higgs super-multiplets
and $\Phi=\{M,Z\}$ collectively denotes all chiral super-multiplets.
The K\"ahler potential $K$ is a real gauge-invariant function. 
The super-potential $W$ and the gauge kinetic function $f(\Phi)$, here referred to the $\SU(3)_c$ color group alone, 
are gauge-invariant analytic functions. We assume a constant gauge kinetic function
\be\label{kinef}
f=\frac{1}{g_{\rm s}^2}-i\frac{\theta_\mathrm{QCD}}{8\pi^2},
\ee
with $\theta_\mathrm{QCD}=0$ having assumed CP invariance.\footnote{We normalize the field strength as $W^a=2 W_B^a$, where 
$W_B^a$ is according to the definition in~\cite{Wess:1992cp}. It follows that $\int d^2\theta \, W^a W^a=-2 F^a_{\mu\nu}F^{a\mu\nu}+2 i
F^a_{\mu\nu} {\tilde F}^{a\mu\nu}$, where ${\tilde F}^{a\mu\nu}=\epsilon^{\mu\nu\rho\sigma}F^a_{\rho\sigma}/2$.
From eq.s~(\ref{susylag}) and (\ref{kinef}) we get $\frac{1}{16} \int d^2\theta~ f(\Phi)  W^a W^a+\hbox{h.c.}=- F^a_{\mu\nu}F^{a\mu\nu}/4g_{\rm s}^2+\theta_\mathrm{QCD} F^a_{\mu\nu} {\tilde F}^{a\mu\nu}/32\pi^2$.}
The super-potential $W$ reads:
\be
W(\Phi)=U^c_i  Y^u_{ij}(Z) Q_j~H_u  +  D^c_i Y^d_{ij}(Z) Q_j~H_d + \cdots
\ee
where $\cdots$ stand for the leptonic and Higgs contributions.

The conditions 1., 2., 3. and 4.~are satisfied assuming that the theory is invariant under
transformations of the type:
\begin{align}
\label{trlambda}
Q_i\to \Lambda^{-k_{Q_i}} Q_i,\qquad U^c_i\to \Lambda^{-k_{U^c_i}} U^c_i,\qquad
D^c_i\to \Lambda^{-k_{D^c_i}} D^c_i,\qquad 
Z_a\to \Lambda^{k_{Z_a}} Z_a, 
\end{align}
where $\Lambda$ is a chiral super-multiplet and $k_{\Phi_i}$ are real weights, here assumed to be integers, with $k_{Z_a}\ge 0$ and
\be
\label{null}
\sum_i\left(k_{Q_i}+k_{U^c_i}+k_{H_u}\right)=\sum_i\left(k_{Q_i}+k_{D^c_i}+k_{H_d}\right)=0.
\ee
In principle the transformations of eq.~(\ref{trlambda}) can be either global or local. We will argue that the interesting case is the one where it is realized by gauge transformations.
Invariance of the superpotential requires:
\begin{eqnsystem}{sys:invariance}
Y^u_{ij}(\Lambda^{k_{Z_1}}Z_1, \ldots, \Lambda^{k_{Z_N}}Z_N)=&~\Lambda^{k_{Q_j}+k_{U^c_i}+k_{H_u}}Y^u_{ij}(Z_1 \ldots ,Z_N),\\
Y^d_{ij}(\Lambda^{k_{Z_1}}Z_1, \ldots, \Lambda^{k_{Z_N}}Z_N)=&~\Lambda^{k_{Q_j}+k_{D^c_i}+k_{H_d}}Y^d_{ij}(Z_1, \ldots ,Z_N),
\end{eqnsystem}
which reproduces condition 2. In the absence of singularities the Yukawa couplings $Y^{u,d}_{ij}(Z)$ can be 
Taylor-expanded in the variables $Z$. 
Only a finite number of terms can satisfy the previous equations, for a given set of weights $k_{D^c_i},k_{U^c_i},k_{Q_j},k_{H_d},k_{H_u}$, implying that $Y^{u,d}_{ij}(Z)$ are polynomials.
As already observed, the property in eq.~(\ref{sys:invariance}) implies that the determinants of $Y^{u,d}_{ij}(Z)$ satisfy
condition 3. Finally, the last condition is a consequence of eq.~(\ref{null}).

To discuss the invariance of the K\"ahler potential, we need to specify the transformations of eq.~(\ref{trlambda}). 
We ask these transformations to realize a gauge group $\Gamma$, continuous or discrete.
Moreover, we allow for this realization to be either linear or non-linear.

If the gauge transformations are linear, the chiral super-multiplet $\Lambda$ depends only on the element $\gamma$ of the group $\Gamma$: $\Lambda=\Lambda(\gamma)$. Moreover, the chiral multiplets $Z$ are elementary.
Collecting all weights in a diagonal matrix 
$k={\diag}(-k_{M_i},k_{Z_a})$, 
the fields transform as
\be
\label{t1}
\Phi\to\Lambda^k (\gamma)\Phi.
\ee
The invariance of the K\"ahler potential is guaranteed by the choice
\be\nn
K(e^{2V} \Phi,\Phi^\dagger)=\Phi^\dagger e^{2V} \Phi.
\ee
The vector super-multiplet $V$ should contain an abelian component that in general is a mixture of SM and possible BSM vectors,
$V=V_{\rm SM}+k V_{\rm BSM}$, where $k$ is the weight matrix.
We will need flavor symmetries different from the SM gauge symmetries, so that the vector super-multiplet transforms as
\be
\label{t2}
V_{\rm SM}\to V_{\rm SM},\qquad
2V_{\rm BSM}\to 2V_{\rm BSM}-\ln\Lambda(\gamma)-\ln\bar\Lambda(\gamma).
\ee
The full action is invariant under eq.~(\ref{trlambda}).

\medskip

If the group $\Gamma$ is non-linearly realized, there will be an elementary chiral multiplet $\tau$ undergoing a 
non-linear transformation, $\tau\to \gamma\tau$. In this case, both $\Lambda$ and $Z$ are composite chiral super-multiplets
depending on $\tau$: $\Lambda=\Lambda(\gamma,\tau)$, $Z=Z(\tau)$.
The gauge transformations should obey the consistency conditions
\be\nn
(\gamma_1\gamma_2)\tau=\gamma_1(\gamma_2\tau),\qquad
\Lambda(\gamma_1\gamma_2,\tau)=\Lambda(\gamma_1,\gamma_2\tau)\Lambda(\gamma_2,\tau),
\ee 
for any pairs of group elements $\gamma_{1,2}$. Now the independent chiral multiplets
are the matter multiplets $M_i$ and $\tau$, 
and the realization of the group $\Gamma$ over the field space and the composite operators $Z$ reads:
\begin{align}
\label{trlambdanl}
\left\{
\begin{array}{rl}
\tau\to&\gamma\tau , \\
M\to& \Lambda(\gamma,\tau)^{-k_M} M,\\
Z_a(\tau)\to& Z_a(\gamma\tau)=\Lambda(\gamma,\tau)^{k_{Z_a}} Z_a(\tau).
\end{array}
\right.
\end{align}
A minimal form of the K\"ahler potential is
\be
K=M^\dagger e^{2V} M+K(\tau,\tau^\dagger).
\ee
As in the linear case, the vector super-multiplet $V$ should contain an abelian component $V_{\rm BSM}(\tau,\tau^\dagger)$: $V=V_{\rm SM}+kV_{\rm BSM}(\tau,\tau^\dagger)$
transforming as
\be
2V_{\rm BSM}(\tau,\tau^\dagger)\to 2V_{\rm BSM}(\tau,\tau^\dagger)-\ln\Lambda(\gamma,\tau)-\ln\bar\Lambda(\gamma,\tau^\dagger).
\ee
Finally, the contribution depending on the superfield $\tau$ alone, $K(\tau,\tau^\dagger)$, should be
invariant under $\tau\to\gamma\tau$ up to K\"ahler transformations:
\be
K(\tau,\tau^\dagger)\to K(\tau,\tau^\dagger)+f(\gamma,\tau)+\bar f(\gamma,\tau^\dagger).
\ee

\subsection{Possible chiral symmetries: U(1) and modular}\label{subex}
A minimal example of a linearly realized symmetry is a spontaneously broken abelian flavor symmetry,
$\Gamma=\U(1)$. In this case we can write
\be\nn
\Lambda=e^{ C},
\ee
where $C$ is a chiral superfield. The chiral super-multiplets $Z_a$ carry positive charges $k_{Z_a}$
and their VEVs spontaneously break CP and U(1). The vector super-multiplet $V_{\rm BSM}$ transforms as
\be
2V_{\rm BSM}\to 2V_{\rm BSM}-C-\bar C.
\ee

\medskip

An example of non-linearly realized symmetry is exhibited by modular-invariant theories, 
where $\Gamma$ is the discrete $\SL(2,\mathbb{Z})$ group and the modulus $\tau$ takes values
in the upper half plane, $y\equiv-i \tau+i\bar\tau>0$. 
For any element of the group $\SL(2,\mathbb{Z})$
\be
\label{sl2z}
\gamma=
\left(
\begin{array}{cc}
a&b\\ c&d
\end{array}
\right)\qquad\hbox{with integer $a,b,c,d$ and}\qquad ad-bc=1\,,
\ee
we define
\be
\tau\to\gamma\tau=\frac{a\tau+b}{c\tau+d},\qquad \Lambda(\gamma,\tau)\equiv c\tau+d.
\ee
Matter fields transform as
\be
M_i\to (c\tau+d)^{- k_{M_i}} M_i\,,
\ee
and the composite operators $Z$ are modular forms of weight $k_{Z_a}$:
\be\nn
Z_a(\gamma\tau)=(c\tau+d)^{k_{Z_a}} Z_a(\tau).
\ee
The $M$-independent part of the K\"ahler potential is
\be
K(\tau,\tau^\dagger)=-h^2\ln y\qquad \hbox{with}\qquad h^2>0,
\ee
where $h$ has mass-dimension one.
$K(\tau,\tau^\dagger)$ is invariant up to a K\"ahler transformation since $y$ transforms as
$ y\to y/|c\tau+d|^2$.
The vector super-multiplet $V_{\rm BSM}$ is built out of $\tau$,
$2V_{BSM}=\ln y$,
so that the minimal K\"ahler potential in the matter sector reads:
\be
M^\dagger e^{-2k_M V_{\rm BSM}} M=M^\dagger y^{-k_M} M.
\ee
\subsection{Anomalies}
\label{anomalies}
We regarded the group $\Gamma$ as a gauge group, and the transformations of eq.~(\ref{trlambda}) as local.
First of all, when eq.~(\ref{trlambda}) provides a non-linear realization of $\Gamma$, the parameter $\Lambda(\gamma,\tau)$
is always local, since the field $\tau$ depends on the space-time coordinates. 
In particular, at the level of the fermionic components of the chiral superfields, the laws in eq.~(\ref{trlambda}) induce local chiral transformations.
When dealing with a linearly realized group $\Gamma$, we could in principle consider global transformations.
Even if non-anomalous, global symmetries are believed to be broken by gravitational effects.
Here we will not consider global non-anomalous symmetries. 
Finally, if continuous and anomalous, global symmetries provide a specific implementation
of the axion paradigm, a scenario alternative  to the one explored here. 
Therefore, we are led to
regard the transformations in eq.~(\ref{trlambda}) as local ones, subject to the requirement of the absence of gauge anomalies.

In particular, a necessary requirement is the absence of anomalies involving the SM field strengths.
Since both $\tau$ (if present) and $Z_a$ are invariant
under the SM gauge group, the relevant diagrams involve only loops of the matter multiplets $M_i=\{U^c_i,D^c_i,E^c_i,Q_i,L_i,H_{u,d}\}$. 
Extending to the lepton sector the transformations of eq.~(\ref{trlambda}),
we consider the variation in the path integral measure ${\cal D}M$ induced by transformations
\be\label{matter}
M_i \to M_i'=\Lambda(\gamma,\tau)^{- k_{M_i}} M_i.
\ee
The ${\cal D}M'=J~{\cal D}M$ variation induces a Jacobian $J$ given by~\cite{Konishi:1985tu,Arkani-Hamed:1997qui}:
\be
\label{Jac}
\ln J=-\frac{i}{64\pi^2}\int d^4 x \, d^2\theta~\sum_M\left[ T(M)k_M \right] W^a W^a\, \ln\Lambda\,,
\ee
where $T(M)$ is the Dynkin index of the various $M$ representations.
We use the normalization $T(N)=1/2$ for the fundamental of $\SU(N)$, and
a compact notation where labels of the individual factors of the gauge group $\SU(3)_c\otimes\SU(2)_L\otimes\U(1)_Y$ 
are not displayed.
Notice that, since $\Lambda$ is complex, this Jacobian accounts for both phase and scale transformations.
The result of eq.~(\ref{Jac}) can be rephrased by saying that the transformation in eq.~(\ref{matter}) induces the following change of the gauge kinetic functions:\footnote{The transformation of eq.~(\ref{matter}) modifies the argument of the determinant of the quark mass matrix
$m$ as $\arg\det m'=\arg\det m+\sum_k(k_{U^c_k}+k_{D^c_k}+2 k_{Q_k})\arg\Lambda$. 
At the same time, the parameter $\theta=-8\pi^2{\Im}f$ is modified into $\theta'=-8\pi^2{\Im}f'=\theta-\sum_k(k_{U^c_k}+k_{D^c_k}+2 k_{Q_k})\arg\Lambda$ 
and the combination $\bar\theta=\theta+\arg\det m$ remains invariant. Furthermore, a chiral rotation of gluinos 
modifies $\theta$ and the redefinition-invariant combination is $\bar\theta=\theta+\arg\det m+C({\rm SU(3)})\arg m_g$,
where $C({\rm SU(3)})=3$ is the Casimir operator and $m_g$ is the gluino mass.}
\be\label{gkfs}
f \to f+\frac{1}{4\pi^2}~\sum_M\left[ T(M)k_M \right]\ln\Lambda.
\ee
Unless we modify our choice of gauge kinetic function to compensate for the shift in eq.~(\ref{gkfs}), we should impose
the conditions:
\be\label{anofree}
\sum_M  T(M)k_M =0,
\ee
which, in the three channels  $\SU(3)_c^2$, $\SU(2)^2_L$ and ${\rm U(1)}_Y^2$, read:
\begin{eqnsystem}{sys:ano1}
&&\sum_i (2k_{Q_i}+k_{U^c_i}+k_{D^c_i})=0 , \label{eq:ano13} \\
&&\sum_i (3k_{Q_i}+k_{L_i})+(k_{H_u}+k_{H_d})=0,\\
&&\sum_i (k_{Q_i}+8k_{U^c_i}+2k_{D^c_i}+3k_{L_i}+6k_{E^c_i})+3(k_{H_u}+k_{H_d})=0.
\end{eqnsystem}
We see that the conditions in eq.~(\ref{null}) to get a real determinant of the quark mass matrix are compatible with 
the equations (\ref{sys:ano1}). Together they imply $k_{H_u}+k_{H_d}=0$. Conversely, if we make use
of the equivalence relations in eq.~(\ref{shift1}) to set $k_{H_u}+k_{H_d}=0$, we see that the absence of $\SU(3)_c$ anomalies
is equivalent to the requirement of a real determinant of the quark mass matrix.

\section{Extensions of Nelson-Barr models}\label{HQ}
The general framework described in the previous section can be successfully applied to solve the strong CP problem in multiple classes of models. 
Realizations based on a modular symmetry will be discussed in section~\ref{cmod}.
We here show how the general framework also includes as a special case the Nelson-Barr class of models \cite{Nelson:1983zb,Barr:1984qx}, 
which solves the strong CP problem relying on extensions of the SM with vector-like heavy quarks.

\smallskip

New chiral quarks gaining mass through Yukawa couplings would affect Higgs production and decay 
and are experimentally ruled out. In contrast, gauge-invariant mass terms for vector-like quarks are easily compatible
with existing bounds, while offering a variety of signatures for future experiments, ranging from violations
of unitarity of the CKM matrix to flavor-changing neutral currents and new collider final states.
Heavy vector-like quarks are expected in several extensions of the SM, such as 
models with extra dimensions in flat or warped geometry, grand unified theories and
composite Higgs models. 
Finally, vector-like quarks, which do not contribute to pure SM gauge anomalies,
can also be useful in canceling anomalies related to gauge extensions of the SM,
as is the case of our framework.
For a recent review on vector-like electroweak-singlet quarks, see~\cite{Alves:2023ufm}.

Focussing on the down-quark sector,
a possible model extends the three generations of quarks $Q_i\sim (3,2,+1/6)$, $D^c_i\sim(\bar 3,1,+1/3)$ with $i=\{1,2,3\}$
by adding $P$ vector-like electroweak-singlet down quarks 
\beq 
D'_\alpha\sim (3,1,-1/3),\qquad
D'^c_\alpha\sim (\bar 3,1,+1/3),\qquad\hbox{with} \qquad\alpha=\{1,...,P\}.\eeq 
The theory is described by\footnote{In what follows, we write the mass matrices in the left-right convention.}
\begin{eqnsystem}{sys:WHQ1}
K_{\rm UV}&=&\sum_\Phi \Phi^\dagger e^{2 V} \Phi=Q^\dagger Q+ D^{c\dagger} D^c+D'^\dagger D'+D'^{c\dagger}D'^c+ H_d^\dagger H_d+\cdots \\
W_{\rm UV}&=&~y^d_{ij} ~Q_i D^c_j H_d+ y'^d_{i\beta} ~Q_i   D'^c_\beta H_d+
N_{\alpha i}~ D'_\alpha~ D^c_i +M_{\alpha\beta} ~ D'_\alpha D'^c_\beta+ \cdots\\
f_{\rm UV}&=& f_0.
\end{eqnsystem}
The resulting down-quark mass matrix can be written as
\beq \label{eq:Mfull}
{\cal M}_d =\bordermatrix{ & D^c_j & D'^c_\beta \cr
 Q_i & m_{ij} & n_{i\beta}\cr
 D'_\alpha & N_{\alpha j} & M_{\alpha\beta}}.\eeq
The up-quark sector can contribute to $K_{\rm UV}$ and $W_{\rm UV}$ with terms similar to those shown in eq.~(\ref{sys:WHQ1})
and will not be displayed here.
The mass matrices $m_{ij}=y^d_{ij} v_d$ and $n_{i\beta}=y'^d_{i\beta} v_d$ arise after electroweak symmetry breaking, 
while $N_{\alpha j}$ and $M_{\alpha\beta}$ are not constrained by the electroweak symmetry. 
The matrices $m_{ij}$, $n_{i\beta}$, $N_{\alpha j}$ and $M_{\alpha\beta}$ depend on dynamical variables $Z$ and, by CP invariance, satisfy
$\bar m(Z)=m(\bar Z)$, etc.
The theory is invariant under the local chiral transformations\footnote{We assume invariant Higgs multiplets.
The discussion can be easily extended to the case where the Higgs multiplets transform nontrivially.}
\beq
\label{chiral}
Q\to \Lambda^{-k_Q}Q, \quad
{D^c}\to \Lambda^{-k_{D^c}} D^c,\quad
D'\to \Lambda^{-k_{D'}} D', \quad
D'^c\to\Lambda^{-k_{D'^c}} D'^c,\quad
Z \to \Lambda^{k_Z}Z,
\eeq
provided the matrices in eq.~(\ref{sys:WHQ1}), through their dependence on the fields $Z$, transform as
\beq
\label{TM}
m\to \Lambda^{k_Q}m \Lambda^{k_{D^c}} , \qquad
n\to \Lambda^{k_Q}n \Lambda^{k_{D'^c}}, \qquad
N\to \Lambda^{k_{D'}}N \Lambda^{k_{D^c}},\qquad
M\to \Lambda^{k_{D'}}M \Lambda^{k_{D'^c}}.
\eeq
In this section, we  concentrate on
the case of a linearly realized symmetry such that,
to describe quark masses, we can safely set to zero the vector superfield $V$. 
This is no longer true if the symmetry is non-linearly realized. 
In this case, the vector multiplet $V$ contains a composite component, depending on the field $\tau$, that can contribute to the mass matrices. 
Below we briefly discuss such a case.

When the transformations 
are non-linearly realized, the vector multiplet $V$ depends on the chiral super-multiplets $(Z,\bar Z)$, and its VEV can be different from zero. 
It is no longer correct to set $V=0$, as we do in the case of linearly realized symmetry.
A non-zero VEV of $V$ modifies the K\"ahler potential of eq.~(\ref{sys:WHQ1}), which now reads
\begin{align}
K_{\rm UV}=\Phi^\dagger e^{2 \langle V\rangle} \Phi=\Phi^\dagger \xi_\Phi^\dagger \xi_\Phi \Phi,
\end{align}
where we have introduced spurions $\xi_\Phi$. 
The theory is formally invariant under the local transformations of eq.~(\ref{chiral}), if the spurions $\xi_\Phi$ transform as: $\xi'_\Phi=\xi_\Phi\Lambda^{k_\Phi}$. It follows that the new variables
$\hat\Phi=\xi_\Phi \Phi$ are invariant and have a canonical normalization. If we express the super-potential $W$ in terms of the hatted fields,
we obtain an expression identical to the one in eq.~(\ref{sys:WHQ1}), 
where the matrices $m$, $n$, $N$ and $M$ have to be replaced by hatted matrices:
\be
\label{eq:hatted}
\hat m=\xi^{-1}_Q m \xi^{-1}_{D^c}\,, \qquad
\hat n=\xi^{-1}_Q n \xi^{-1}_{D'^c}\,, \qquad
\hat N=\xi^{-1}_{D'} N \xi^{-1}_{D^c}\,,\qquad 
\hat M=\xi^{-1}_{D'} M \xi^{-1}_{D'^c}\,,
\ee
and similarly for the up sector.
These matrices are also formally invariant under the local transformations of eq.~(\ref{chiral}). 

\medskip

The condition for the absence of mixed chiral/$\SU(3)_c^2$ anomalies is
\be
\label{noan}
\sum_i(k_{Q_i}+k_{D^c_i})+\sum_\alpha(k_{D'_\alpha}+k_{D'^c_\alpha})+ \cdots=0,
\ee
where $\cdots$ stand for the contribution of the up-quark sector and possible additional vector-like quarks.
The determinant of the down-quark mass matrix ${\cal M}_d$
 is a homogeneous polynomial
in the variables $Z$ of degree
\be
\label{dd}
d_d=\sum_i(k_{Q_i}+k_{D^c_i})+\sum_\alpha(k_{D'_\alpha}+k_{D'^c_\alpha}).
\ee
Similarly, the determinant in the up-quark sector is a homogeneous polynomial
in the variables $Z$ of degree $d_u$, whose expression is analogous to eq.~(\ref{dd}).
Choosing weights such that $d_d=d_u=0$, the determinant of the overall quark mass matrix is real and
the condition for the cancellation of mixed chiral/$\SU(3)_c^2$ anomalies of eq.~(\ref{noan}) is satisfied.

This is enough to show that the full theory solves the strong CP problem,
as long as the observed quark masses and mixings, including the CKM phase, are successfully reproduced.
To check this, we next compute the effective mass matrix of light quarks.
We will also explicitly show how the strong CP problem is solved from the point of view of the low-energy EFT.
This discussion requires clarifying some properties of the EFT, 
as in general the chiral symmetry can become anomalous when restricted to light fields,
such that anomaly cancellation also involves the gauge kinetic functions.
We provide a pedagogical technical computation of this issue, 
while emphasising that this step is not necessary, as the key physical result $\bar\theta=0$ is
simply explicit in the full theory.

\subsection{The mass matrix of light quarks}
Assuming $|M|,|N|\gg |m|,|n|$, we can describe the low-energy properties of this theory by integrating out the heavy states.
We work in the broken phase, where both the electroweak symmetry and CP are spontaneously broken
and $m$, $n$, $N$, $M$ should be understood as numerical matrices.
From eq.~(\ref{sys:WHQ1}) we see that $D'$ and the combinations $N_{\alpha i} D^c_i+ M_{\alpha\beta} D'^c_\beta$ are heavy. 
We define a new orthonormal set of states $(d^c,h^c)$, through the unitary transformation:
\be\nn
\left(
\begin{array}{c}
d^c\\
h^c
\end{array}
\right)=\left(
\begin{array}{cc}
\left[ \One +N^\dagger (M M^\dagger)^{-1} N\right]^{-1/2}&-\left[ \One +N^\dagger (M M^\dagger)^{-1} N\right]^{-1/2}N^\dagger M^{-1\dagger}\\
(M M^\dagger+N N^\dagger)^{-1/2} N & (M M^\dagger+N N^\dagger)^{-1/2} M
\end{array}
\right)\left(
\begin{array}{c}
D^c\\
D'^c
\end{array}
\right).
\ee
After electroweak and CP breaking, the theory, expressed in terms of $Q=(U,D)$, $D'$, and $(d^c,h^c)$, is described by
\begin{align}
K_{\rm UV}=&~D^\dagger D+ d^{c\dagger} d^c+D'^\dagger D'+h^{c\dagger}h^c+\cdots\\
W_{\rm UV}=&~D^T~(m-n M^{-1}N) [ \One +N^\dagger (M M^\dagger)^{-1} N]^{-1/2} ~ d^c\nn\\
+&~D^T~(mN^\dagger+n M^\dagger) (M M^\dagger+N N^\dagger)^{-1/2}   ~ h^c
+D'^T~ (M M^\dagger+N N^\dagger)^{1/2}~ h^c+\cdots
\end{align}
At the tree level, the low-energy effective theory is obtained by eliminating the heavy sector $(h^c,D')$ by means of the static equations of motion:
\begin{align}
h^c=0,\qquad 
D^T~(m~N^\dagger+n~ M^\dagger) (M M^\dagger+N N^\dagger)^{-1/2}
+D'^T~ (M M^\dagger+N N^\dagger)^{1/2}=0.
\end{align}
Substituting these equations in the canonical K\"ahler potential $K_{\rm UV}$ and in the superpotential $W_{\rm UV}$ gives
\begin{eqnsystem}{sys:WEFT}
\hat K_{\rm IR}&=&D^\dagger[ \One +(\bar m N^T+\bar n  M^T)(M M^\dagger+N N^\dagger)^{-2~T}(\bar N m^T+\bar M  n^T)] D+d^{c\dagger} d^c+\cdots\\
\hat W_{\rm IR}&=&D^T~(m-n M^{-1}N) [ \One +N^\dagger (M M^\dagger)^{-1} N]^{-1/2} ~ d^c+  \cdots
\end{eqnsystem}
The correction to the $D^\dagger D$ kinetic term is suppressed in the $|M|,|N|\gg |m|,|n|$ limit and could be omitted
restoring explicit $\SU(2)_L$ invariance.
Furthermore, $\hat W_{\rm IR} $ is not explicitly holomorphic;
supersymmetry is made manifest by describing the same low-energy theory
in an equivalent way that moves the non-holomorphic part to the $d^{c\dagger}d^c$ kinetic term:
\begin{eqnsystem}{sys:KWIR}
K_{\rm IR}&=&~D^\dagger[ \One  + {\cal O}(m,n/M,N)^2
]  D
+d^{c\dagger} [ \One +N^\dagger (M M^\dagger)^{-1} N] d^c +\cdots \\
W_{\rm IR}&=& D^T~ (m-n M^{-1}N)~ d^c+ \cdots ,
\end{eqnsystem}
where now $m$, $n$, $N$, $M$ are understood as operators depending on the chiral super-multiplets $H_d$ and $Z$.
The physical down-quark mass matrix is\footnote{At leading order in $m,n$ the result
can be written in an equivalent way that renders manifest the symmetry under $m\leftrightarrow n$ and 
$M\leftrightarrow N$~\cite{Alves:2023ufm},
$
m_d m_d^\dagger = m\,m^\dagger+n\,n^\dagger-(m\,N^\dagger+n\,M^\dagger)(MM^\dagger+NN^\dagger)^{-1}(Nm^\dagger +Mn^\dagger)$.}
%
%
\begin{align}
\label{md}
m_d=&~( \One +(m~N^\dagger+n~ M^\dagger)(M M^\dagger+N N^\dagger)^{-2}(N~m^\dagger+M~ n^\dagger))^{-1/2}\times\nn\\
&~\times(m-n M^{-1}N)( \One +N^\dagger (M M^\dagger)^{-1} N)^{-1/2}.
\end{align}
The matrix $m_d$ has a holomorphic part, $m^{\rm hol}_d=m-n M^{-1}N$, which 
contributes to the $\bar\theta_{\rm IR}$ parameter. 
The non-holomorphic part of $m_d$ does not contribute to $\bar\theta_{\rm IR}$, since it consists of products of hermitian matrices. It can contribute to the CKM phase.
The determinant  $\det(m-n M^{-1}N)$ is no longer a constant, but a polynomial in $Z$ of degree $\sum_i(k_{Q_i}+k_{D^c_i})\ne0$ and 
its phase will be non-vanishing.

The results derived here in the case of a linearly realized symmetry 
remain valid also in the case of a non-linearly realized symmetry, 
provided we replace everywhere fields and matrices with their hatted counterparts given in eq.~\eqref{eq:hatted}.
In particular,
as we see from eq.~(\ref{md}), the $\xi_\Phi$ factors will provide an extra contribution to the low-energy quark mass matrix.

\subsection{Chiral anomalies in the low-energy EFT}

The definition of the low-energy effective theory is completed by the gauge kinetic function. In the full theory,
this is given by a real constant $f_{\rm UV}$. In the low-energy theory, we expect it to acquire a dependence
on the $Z$ fields. 
We parameterise the result as
\be
\label{fIR}
f_{\rm IR}(Z)=f_0(Z)-\frac{1}{8\pi^2} \ln\det M(Z)+\cdots
\ee
where $f_0(Z)$ is a generic function, and now show that it is simply a constant.
To determine $f_0(Z)$ we first ask that the full and low-energy theory deliver
the same parameter $\bar\theta$, order-by-order in perturbation theory. At the lowest order, we have
\begin{align}
\bar\theta_{\rm UV}=&~\arg\det {\cal M}_d+\cdots =\arg\det M+\arg\det (m-n M^{-1}N)+\cdots\nn\\
\bar\theta_{\rm IR}=&-8\pi^2 {\Im} f_{IR}+\arg\det m_d+\cdots  \\
=&
-8\pi^2 {\Im} f_{0}(Z)+{\Im}\ln\det M+\arg\det (m-n M^{-1}N)+\cdots
\end{align}
By requiring $\bar\theta_{\rm IR}=\bar\theta_{\rm UV}$ we get 
\be
\label{im}
{\Im} f_{0}(Z)=0.
\ee
Next, we ask for the cancellation of gauge
anomalies, which in the full theory is guaranteed by eq.~(\ref{noan}).
In the path integral we make the change of variables of eq.~(\ref{chiral}). 
As an effect of the anomalous EFT field content,
eq.~(\ref{gkfs}), and of the dependence on $Z$ through $M(Z)$,
the only modification is in the gauge kinetic function $f_{\rm IR}$ that becomes:
\begin{align}
f_{\rm IR} \to &~f_0(\Lambda^{-k_Z} Z)+\frac{1}{8\pi^2} \sum_i(k_{Q_i}+k_{D^c_i})\ln\Lambda-\frac{1}{8\pi^2} \ln\det [\Lambda^{-k_{D'}}M(Z)\Lambda^{-k_{D'^c}}]+ \cdots\\
=&~f_0(\Lambda^{-k_Z} Z)-\frac{1}{8\pi^2}\ln\det M(Z)+\frac{1}{8\pi^2} \left[\sum_i(k_{Q_i}+k_{D^c_i})+\sum_\alpha(k_{D'_\alpha}+k_{D'^c_\alpha})\right]\ln\Lambda+ \cdots \nn \\
=&~f_0(\Lambda^{-k_Z} Z)-\frac{1}{8\pi^2}\ln\det M(Z)+ \cdots
\end{align}
where in the last equality we made use of the condition in eq.~(\ref{noan}).
It follows that the invariance of the theory under chiral rotations of the matter fields requires:
\be
\label{con}
f_0(\Lambda^{-k_Z} Z)=f_0(Z).
\ee
The conditions in eq.~(\ref{im}) and (\ref{con}) imply that $f_0(Z)$ is a real constant. Eq.s~(\ref{sys:KWIR}) and (\ref{fIR}) describe the low-energy theory.

\subsection{An example of an anomalous low-energy EFT} \label{ex1}
We present an explicit example of the previous general discussion.
For simplicity we consider an example with two fermion generations, two species of heavy quarks, and a single multiplet $Z$.
We choose weights 
\beq\label{eq:nexample}
k_{D^c_i}=(0,-1), \quad
k_{Q_i}=(0,-1), \quad
k_{D'_\alpha}=(0,+1), \quad
k_{D'^c_\alpha}=(0,+1), \quad
k_Z=+1.\eeq
Consequently, the down-quark mass matrices are:
\begin{align}\nn
m_{ij}=&~
\left(
\begin{array}{cc}
m^0_{11}&0\\
0&0
\end{array}
\right),
&n_{i\alpha}&=
\left(
\begin{array}{cc}
n^0_{11}&n^0_{12}~ Z\\
0&n^0_{22}
\end{array}
\right),\nn\\
N_{\alpha i}=&~
\left(
\begin{array}{cc}
N^0_{11}&0\\
N^0_{21}~ Z&N^0_{22}
\end{array}
\right),
&M_{\alpha\beta}&=
\left(
\begin{array}{cc}
M^0_{11}&M^0_{12}~ Z\\
M^0_{21}~ Z&M^0_{22}~ Z^2
\end{array}
\right),\nn
\end{align}
where $m^0, n^0, N^0, M^0$ are real parameters.
The weights in eq.\eq{nexample} guarantee that in the full theory  the 
mixed chiral/$\SU(3)_c^2$ anomalies cancel. 
So the determinant of the holomorphic part of the full down-quark mass matrix is real, solving the strong CP problem:
\be
\det{\cal M}^{\rm hol}_d=(n^0_{11} N^0_{11}-m^0_{11} M^0_{11}) n^0_{22} N^0_{22}.
\ee
However, mixed chiral/$\SU(3)_c^2$ anomalies appear in the EFT restricted to light down quarks.
So the determinant of the heavy mass matrix $M$ vanishes for $Z\to 0$,
and the effective mass matrix of light down quarks acquires a non-trivial structure. 
According to eq.~(\ref{md}) the holomorphic part of the effective mass matrix of light down quarks
\beq
m^{\rm hol}_d= m-n M^{-1} N=\frac{1}{\det M^0}
\left(
\begin{array}{cc}
\mu^3_{11}& \mu^3_{12} Z^{-1} \\
\mu^3_{21} Z^{-1} & \mu^3_{22}  Z^{-2} \\
\end{array}
\right),
\eeq
where $\det M^0 = M^0_{11} M^0_{22} - M^0_{12} M^0_{21}$ and
\begin{align}
\mu^3_{11}=&~ m^0_{11} \det M^0 - M^0_{22} n^0_{11} N^0_{11} + M^0_{21} N^0_{11} n^0_{12} + M^0_{12} n^0_{11} N^0_{21} - M^0_{11} n^0_{12} N^0_{21}\,,\nn\\
\mu^3_{12}=&~(M^0_{12} n^0_{11} - M^0_{11} n^0_{12}) N^0_{22}\,,\nn\\
\mu^3_{21}=&~(M^0_{21} N^0_{11} - M^0_{11} N^0_{21} )n^0_{22}\,,\nn\\
\mu^3_{22}=&~-M^0_{11} n^0_{22} N^0_{22}\,.\nn
\end{align}
The entries of $m^{\rm hol}_d$ with negative weights are now filled with negative powers of $Z$. 
The limit $Z=0$ signals the failure of the low-energy approximation due to a vanishing eigenvalue of the heavy-quark mass matrix $M_{\alpha\beta}$. 
The determinant of $m^{\rm hol}_d$
\be\nn
\det m^{\rm hol}_d=\frac{(n^0_{11} N^0_{11}-m^0_{11} M^0_{11}) n^0_{22} N^0_{22}}{Z^2\det M^0}
\ee
acquires a phase from the $Z$ dependence. 
The low-energy effective theory has an anomalous field content, but the anomaly is canceled by the modified
gauge kinetic function
\be\nn
f_{\rm IR}(Z)=f_0-\frac{1}{8\pi^2} \ln\det M(Z)+ \cdots =
f_0-\frac{1}{8\pi^2}\ln(Z^2 \det M^0)\,.
\ee
At the same time, the physical $\bar\theta$ parameter, evaluated in the low-energy theory,
coincides with the one evaluated in the full theory:
\begin{align}
\bar\theta_{\rm IR}=&~-8\pi^2 {\Im} f_{IR}+\arg\det m^{\rm hol}_d\nn\\
=&~{\Im}\ln(Z^2~\det M^0)+
\arg \frac{(n^0_{11} N^0_{11}-m^0_{11} M^0_{11}) n^0_{22} N^0_{22}}{Z^2~\det M^0}\nn\\
=&~
\arg(n^0_{11} N^0_{11}-m^0_{11} M^0_{11}) n^0_{22} N^0_{22} = \bar\theta_{\rm UV}\,.\nn
\end{align}
The phase of the determinant of $m^{\rm hol}_d$ is exactly canceled by imaginary part of
the new gauge kinetic function $f_{\rm IR}$.

\medskip

In this example, there are two quark generations so no CKM phase. In the realistic case of three generations
and several fields $Z$, the CKM phase gets two contributions. The first one is from the $Z$ dependence of 
the holomorphic part of the up and down quark matrices. The second one is from the $Z$ dependence of 
the non-holomorphic parts. Depending on the specific model, one of the two can dominate.
For instance, in the Nelson-Barr framework, the CKM matrix is typically saturated by the contribution from
the non-holomorphic parts, as we show in the next section.

\subsection{Nelson-Barr models as a special case}\label{NB}
The supersymmetric version of the Nelson-Barr solution to the strong CP problem can be viewed as
a particular case of our framework. Nevertheless, its typical realization presents interesting aspects, since the CKM phase arises entirely from a wave function renormalization effect.
In Nelson-Barr models there are two distinct sectors. The first one consists of the SM
quarks, and the second one includes a set of extra particles in a vector-like representation of the SM gauge group. 
As in the general framework analysed here, the theory is CP-invariant and CP is spontaneously broken by a set of 
fields $Z$. In Nelson-Barr models the vanishing of $\bar\theta$ is guaranteed by two assumptions: 
\begin{itemize}
\item[$i)$] The SM electroweak symmetry is broken
only in the pure SM sector. 
\item[$ii)$] CP is broken only in the sector that couples the SM quarks to the extra particles.
\end{itemize}
This typical Nelson-Barr  setup can be reproduced by the theory analysed at the beginning of section \ref{HQ}. 
Focussing on the down-quark sector, to comply with the two conditions $i)$ and $ii)$, we assume that
\begin{itemize}
\item[a)] $n$ vanishes; 
\item[b)] $m$ depends on the Higgs doublet $H_d$ and does not depend on $Z$;
\item[c)] $N$ depends on $Z$ and does not depend on the Higgs doublet $H_d$; 
\item[d)] $M$ does not depend on $Z$ nor on $H_d$. 
\end{itemize}
We can choose a basis where $M_{\alpha\beta}$ is diagonal and positive definite and $N_{\alpha i}=Z_{\alpha i}$ is linear in the $Z$ fields.
We are thereby justifying the Nelson-Barr ansatz by
considering a theory with one U(1) broken by multiple scalars $Z$ that acquire different CP-breaking phases.
We do not discuss how such VEVs can be obtained minimizing a scalar potential.
Up to this issue, the Nelson-Barr super-potential ansatz gets justified by assuming the following weights 
\beq k_{H_d} = k_{Q_i}= k_{D^c_i} = 0,\qquad
k_{D'_\alpha}=-k_{D'^c_\beta} = k_{Z_{\alpha i}}=1
\eeq
so that the chiral symmetry is anomaly free, even when restricted to the light fields.
Together with gauge invariance, this assignment forbids terms of the type $ QD^c Z$, $D'D'^c  Z$, $D' D^c H_d$ $ Q D'^c H_d$ and $D' D'^c H_d$.
The determinant of the mass matrix is a homogeneous polynomial
in the variables $Z_{\alpha i}$ of degree $\sum_i(k_{Q_i}+k_{D^c_i}+k_{H_d})+\sum_\alpha(k_{D'_\alpha}+k_{D'^c_\alpha})=0$, which is a real constant. Indeed,
after CP and electroweak spontaneous breaking, the quark mass matrix reads
\be\nn
{\cal M}_d=\left(
\begin{array}{cc}
m&0\\
Z&M
\end{array}
\right),
\ee
and its determinant,
$\det {\cal M}_d=\det m\det M$,
is real. By specializing the general results of eq.s~(\ref{sys:KWIR}) and (\ref{md}), we get
\begin{eqnsystem}{sys:IRNB}
K_{\rm IR}&=& D^\dagger[ \One +m~Z^T(M^2+Z Z^\dagger)^{-2~T}\bar Z~m^T]  D
+ d^{c\dagger} ( \One +Z^\dagger M^{-2} Z) d^c +\cdots \\
W_{\rm IR}&=& D^T\, m \,d^c+\cdots .
\end{eqnsystem}
In this manifestly supersymmetric description, the contribution to $\bar\theta$ from the down-quark sector is
automatically zero (provided $\det m>0$),
and the CKM phase arises from the transformation needed to put the kinetic terms into a canonical form. Denoting the VEV of $Z$ by $z$, the light quark mass matrix is 
\be\nn
m_d=[ \One +m z^\dagger(M^2+z z^\dagger)^{-2} z m^T]^{-1/2} m~( \One +z^\dagger M^{-2}z)^{-1/2 } .
\ee 
This matrix is hermitian and its determinant is still real,
much as the determinant of the complete matrix ${\cal M}_d$.
In this typical Nelson-Barr model, the chiral symmetries have no QCD anomaly nor in the light sector nor in the full theory.
So there is no compensation between the phases of the determinants in the light and heavy sectors.
The low-energy theory has a non-anomalous field content and its gauge kinetic function is a real constant.
Nevertheless, the light quark mass matrix is complex and can deliver a contribution to the CKM phase. 
As in the general case, the light quark mass matrix depends both on $z$ and on the conjugate variables $\bar z$. Such a dependence can be thought of as associated with non-canonical
kinetic terms arising from the elimination of a heavy sector from the low-energy theory.

\medskip

Thus, the Nelson-Barr models are special cases of a much wider class of models that include heavy vector-like quarks
in their spectrum. In this wider class, low-energy gauge anomalies cancel thanks to the nontrivial properties of the gauge kinetic function. Moreover, both holomorphic and non-holomorphic sectors contribute to the CKM phase.

As anticipated above, a possible issue of such models is realizing a U(1)-invariant super-potential with CP-conserving parameters
such that multiple scalars $z_a$ break the U(1) by acquiring VEVs with different CP-breaking phases.
Furthermore, to write fundamental models rather than effective models,
one needs to introduce extra states able of mediating the desired Yukawas, 
without mediating undesired contributions proportional to negative powers of some $z_a$.
We next consider models where the U(1) is replaced by a modular symmetry, that is automatically broken in the desired way.
The presence of multiple scalars with different CP phases becomes a consequence of 
the mathematics of modular forms.

\section{Higher level modular forms}\label{cmod}
In this section we discuss some concrete realizations of our framework based on the modular invariance anticipated in section \ref{subex}.
Its mathematics automatically provides one key ingredient: a chiral symmetry automatically broken by `scalars' with multiple charges and different CP-violating phases.
Furthermore, the other key ingredient naturally arises in string models, where CP is a symmetry broken by compactifications with a complex structure.
The modular symmetry is how such structure manifests in the effective 4-dimensional theory.
For simplicity, we assume that there is a single modular $\SL(2,\mathbb{Z})$ and thereby a single modulus $\tau$,
and work in the standard basis where CP acts on it as $\tau\to-\tau^\dagger$.
As usual, modular transformations allow to restrict $\tau$ to its fundamental domain: $\Im\tau>0$, $|\tau|\ge 1$ and $|{\rm Re}\,\tau|\le 1/2$.
CP is spontaneously broken when $\tau$ is inside the domain, and away from the imaginary axis, see e.g.~\cite{hep-th/9506143,1901.03251,1905.11970}.

\smallskip

Models of this type, achieving $\bar\theta=0$ and a non-vanishing CKM phase, based on full modular invariance were obtained in~\cite{Feruglio:2023uof}
employing convenient assignments of quark modular weights. 
For example, the choice 
\beq \label{eq:modelfull}
k_{Q_i}=k_{U^c_i}=k_{D^c_i}=\{-6,0,+6\},\qquad k_{H_{u,d}}=0\eeq 
satisfies the conditions for anomaly cancellation and delivers a real determinant for
the quark mass matrices. 
At the same time, the complex nature of $\tau$ allows for an ${\cal O}(1)$ phase in the CKM mixing matrix.%
\footnote{The supersymmetric model based on the traditional $A_4$ flavor symmetry combined with $\U(1)_R$ and $Z_2 \times Z_4^5$ shaping symmetries~\cite{1307.0710}
requires 9 flavons to get a similar structure of the Yukawa matrices.}

While such a minimal realization is welcome, there are reasons to explore non-minimal models, which might exhibit interesting features. 
The solution of~\cite{Feruglio:2023uof} requires large weights of the matter multiplets, not necessarily realized in string theory compactifications (see e.g.~\cite{Ibanez}).
Moreover, due to the large number of free parameters, quark masses, mixing angles and CKM phase cannot be predicted, but just accommodated.
It is plausible that, by varying the matter content of the model, solutions making use of smaller weights and/or less free parameters can be found. 


Here we show that it is possible to formulate models where lower weights of matter fields can be adopted.
The key ingredient to achieve these goals are modular forms of higher levels.

The modular group $\SL(2,\mathbb{Z})$, eq.~(\ref{sl2z}), is generated by the two elements
\be
S=
\left(
\begin{array}{cc}
0&-1\\
1&0
\end{array}
\right),\qquad
T=
\left(
\begin{array}{cc}
1&1\\
0&1
\end{array}
\right),
\ee
satisfying $S^2=(ST)^3=- \One $. If the flavor group coincides with $\SL(2,\mathbb{Z})$, 
we can choose modular forms of level one to be the composite operators $Z$ discussed in the previous sections,
\be
\label{modtrans}
Z^{(k)}(\gamma\tau)=(c\tau+d)^{k} Z^{(k)}(\tau),
\ee
for any element $\gamma$ of $\SL(2,\mathbb{Z})$.
A generic modular form of level one is a weighted-homogeneous polynomial in two 
basic forms of weights four and six, which can be identified with the Eisenstein series $E_{4,6}(\tau)$. Forms of weight smaller than four vanish, except for those of vanishing weight, which are constant.

\medskip

In a more general framework, the flavor group still coincides with $\SL(2,\mathbb{Z})$, 
but the composite operators $Z$ are modular forms of higher {\em levels} $N$. 
Levels are introduced by assuming that the transformation in eq.~(\ref{modtrans}) is a symmetry only under a subgroup of
$\SL(2,\mathbb{Z})$.
The principal congruence subgroup of level $N$, $\Gamma(N)$, is defined as
the subgroup of $\SL(2,\mathbb{Z})$ matrices
\be
\gamma=\left(
\begin{array}{cc}
a&b\\c&d
\end{array}
\right) \qquad  \hbox{that satisfy}\qquad
\gamma\equiv
\left(
\begin{array}{cc}
1&0\\0&1
\end{array}
\right)~{\rm mod}\, N .
\ee
Modular forms of level $N$ and weight $k$, denoted as $Z^{(k)}(\tau)$, are holomorphic functions of $\tau$ transforming as in eq.~(\ref{modtrans})
for any element $\gamma$ of $\Gamma(N)$. As for modular forms of level one, negative-weight forms vanish
and zero-weight forms are constant. 
Modular forms of level $N$ and weight $k$ can be written as linear combinations of a basis of $k+1$ functions $Z^{(k)}_a(\tau)$,
that transform nicely under the full modular group $\SL(2,\mathbb{Z})$.
In technical language, they span a linear space  ${\cal M}_k(\Gamma(N))$ of finite dimension.
Indeed the group $\Gamma(N)$ is a normal subgroup of $\SL(2,\mathbb{Z})$ and the quotient
$\Gamma_N=\SL(2,\mathbb{Z})/\Gamma(N)$ is a finite group.\footnote{For small values of $N$, $\Gamma_N$ is isomorphic to (the double cover of) 
small permutation groups: $\Gamma_2\sim S_3$, $\Gamma_3\sim T'$, $\Gamma_4\sim S_4'$
and $\Gamma_5\sim A_5'$.   The finite $\Gamma_N$ should not be confused with the infinite $\Gamma(N)$.}
The basis of functions $Z^{(k)}_a(\tau)$ in ${\cal M}_k(\Gamma(N))$ can be chosen
such that, for any element $\gamma$ in the full modular group $\SL(2,\mathbb{Z})$, they transform as
\be
\label{yesrho}
Z^{(k)}_a(\gamma\tau)=(c\tau+d)^k \rho(\gamma)_{ab}Z^{(k)}_b(\tau),
\ee
where $\rho(\gamma)_{ab}$ is a unitary representation of the group $\Gamma_N$. 
This transformation generalises eq.~(\ref{modtrans}).
As a result, nontrivial modular forms of positive weight smaller than four exist at higher levels $N>1$.

\medskip

Therefore, modular forms of higher levels make it possible to assign smaller weights to matter fields and, at the same time,
the non-abelian character of their transformation laws can be useful in reducing the number of free parameters.
By making use of forms of higher levels we can build models of two types:
\begin{enumerate}
\item We assume that the flavor symmetry group is one of the principal congruence subgroups $\Gamma(N)$ with higher level $N>1$.
In this case, any non-vanishing entry of the quark mass matrix involves one or more independent parameters, much as in level $N=1$ models, but smaller weights become available.
This will be discussed in sections~\ref{Gamma(2)} and \ref{Gamma(3)}.

\item  We adopt the full modular group $\SL(2,\mathbb{Z})$ as flavor symmetry, 
but we allow matter multiplets and composite operators $Z$ to transform
as in eq.~(\ref{yesrho}), in non-trivial representations of the finite group $\Gamma_N$. 
Physically, this amounts to non-trivially embed flavor into the modular group, 
such that basic representations involve more than one flavor and lower weight $k$.
The non-abelian character of $\Gamma_N$ may result in a reduction of the number of free parameters of the model,
as the action is formed out of singlet contractions invariant under the full modular group. 
For recent reviews on the modular invariance approach 
to the flavor problem~\cite{Feruglio:2017spp} realizing 
this second option, see~\cite{2307.03384,2311.09282}.
\end{enumerate}
String compactifications can lead to the structure 2., with $\Gamma_N$ related e.g.\ to geometric symmetries of orbifold  fixed points.
The structure 1.\ can arise in compactifications with fluxes that break the full modular group~\cite{2311.12425,2405.18813}.


\section{Models with modular symmetry at level $N=2$}\label{N=2}
We start  considering $N=2$. The principal congruence subgroup $\Gamma(2)$ is generated by the two independent elements:
\begin{equation}
\label{eq:gen_gamma2}
G_1=\begin{pmatrix}
1 & 2\\
0 & 1
\end{pmatrix}=T^2,
\qquad \text{and} \qquad 
G_2=\begin{pmatrix}
3 & -2\\
2 & -1
\end{pmatrix}=(T^2S)^2\,.
\end{equation}
Since $-\mathbb{1}\in \Gamma(2)$, the only non-zero modular forms of level $2$ have even weight. For the same reason, the quotient group $\Gamma_2=\SL(2,\mathbb{Z})/\Gamma(2)$ is its own double cover, and it is isomorphic to the non-abelian discrete permutation group $S_3$. 
The finite group $\Gamma_2$  is generated by elements $\tS$ and $\tT$ that satisfy
\begin{equation}
\label{eq:gen_s3}
\tS^2=(\tS \tT)^3=\tT^2= \mathbb{1}\,,
\end{equation}
and it admits three irreducible representations: two singlets $\mathbf{1}_0$, $\mathbf{1}_1$, and the doublet $\mathbf{2}$. 
The completely neutral trivial singlet is $\mathbf{1}_0$.
See appendix~\ref{g2} for the group properties of $\Gamma_2$, 
including tensor products of its representations 
and the related Clebsch-Gordan coefficients.

There are $k/2+1$ linearly independent modular forms of level 2 and weight $k$, forming a space denoted as ${\cal M}_k(\Gamma(2))$.
For $k=0$ the only form is a constant.
At lowest non-trivial weight $k=2$ there are two linearly independent modular forms: $Z^{(2)}_{a}$ with $a=\{1,2\}$.
They can be built from the Dedekind eta function $\eta(\tau)$~\cite{1803.10391}:
\begin{eqnsystem}{sys:Z2a}
 Z^{(2)}_{1} &=& \frac{2i}{\pi} \left[\frac{\eta'(\tau/2)}{\eta(\tau/2)} + 
 \frac{\eta'((\tau+1)/2)}{\eta((\tau+1)/2)} - 8\frac{\eta'(2\tau)}{\eta(2\tau)}  \right]\,, \\
 \qquad
 Z^{(2)}_{2} &=& \frac{2\sqrt{3}i}{\pi} \left[\frac{\eta'(\tau/2)}{\eta(\tau/2)} -  \frac{\eta'((\tau+1)/2)}{\eta((\tau+1)/2)} \right],
\end{eqnsystem}
normalized such that the $q$-expansion of $Z^{(2)}_1$ starts with 1, see appendix~\ref{g2}.

The transformation properties of $\eta(\tau)$ imply that they transform as in eq.~(\ref{modtrans}) with $k=2$ under the group $\Gamma(2)$. 
Modular forms $Z^{(k)}$ of higher weights $k \ge 4$ are homogeneous polynomials in $Z^{(2)}_{1,2}$. 
For instance, the 3 linearly independent modular forms of weight $k=4$ are
\begin{equation}
 Z^{(4)}_a =\left\{ {Z^{(2)}_{2}}^2 - {Z^{(2)}_{1}}^2\,,~
 2 Z^{(2)}_{1} Z^{(2)}_{2}\,,~
 {Z^{(2)}_{1}}^2 + {Z^{(2)}_{2}}^2 \right\}\,.
 \label{eq:N2k4}
\end{equation}
Under the full $\SL(2,\mathbb{Z})$, the two forms $Z^{(2)}_{1,2}$ 
fill a doublet $Z^{(2)} = (Z^{(2)}_{1}, Z^{(2)}_{2})^T$,
whereas the three forms $Z^{(4)}_{1,2,3}$ arrange themselves into 
a doublet $Z^{(4)} = (Z^{(4)}_{1}, Z^{(4)}_{2})^T$
and an invariant singlet $Z^{(4)}_3$. They transform as
\begin{align}
 Z^{(2)}_{}(\gamma\tau) &=(c\tau+d)^2\,\rho_\mathbf{2}(\gamma)\,Z^{(2)}_{}(\tau)\,, \\
 Z^{(4)}_{}(\gamma\tau) &=(c\tau+d)^4\,\rho_\mathbf{2}(\gamma)\,Z^{(4)}_{}(\tau)\,, \\
 Z^{(4)}_3(\gamma\tau) &= (c\tau+d)^4\,Z^{(4)}_3(\tau)\,,
\end{align}
where the matrix
$\rho_\mathbf{2}$ is the unitary representation $\mathbf{2}$ of $\Gamma_2 \cong S_3$, see appendix~\ref{g2}.

\begin{table}[t]
\centering
\renewcommand{\arraystretch}{1.3}
\resizebox{1.0\textwidth}{!}{\begin{tabular}{ l | c c c c c c c c}
Observable
 & $\displaystyle \frac{m_u/m_c}{10^{-3}}$ & $\displaystyle\frac{m_c/m_t}{10^{-3}}$ & $\displaystyle\frac{m_d/m_s}{10^{-2}}$ & $\displaystyle\frac{m_s/m_b}{10^{-2}}$ & $\displaystyle\frac{\sin^2\theta_{12}}{10^{-2}}$ & $\displaystyle\frac{\sin^2\theta_{13}}{10^{-5}}$ & $\displaystyle\frac{\sin^2\theta_{23}}{10^{-3}}$ & $\displaystyle\frac{\delta_{\rm CKM}}{\pi}$ \\[0.2cm]
\hline 
Central value & $2.04$ & $2.68$ & $5.05$ & $1.37$ & $5.08$ & $0.99$ & $1.28$ & 0.385 \\
Uncertainty & 1.27 & 0.25 & 1.24 & 0.15 & 0.06 & 0.31 & 0.48 & 0.034 
\end{tabular}}
\caption{\em Central values of and $1\sigma$ uncertainties 
on the quark mass ratios, mixing angles and the CKM phase,
renormalized at the GUT scale $2~10^{16}\GeV$ for $\tan\beta=5$~\cite{1306.6879}.}
\label{tab:data}
\end{table}

\subsection{A model with $\Gamma(2)$ modular symmetry}\label{Gamma(2)}
We here assume that $\Gamma(2)$ is the flavor symmetry group, and propose a supersymmetric
model that only
includes the three  generations of SM quarks (i.e.~no heavy quarks),
$Q_{1,2,3}$, $D^c_{1,2,3}$ and $U^c_{1,2,3}$ and the MSSM Higgs doublets $H_{u,d}$.
We assume the minimal weights
\beq \label{eq:model1}
k_{Q_i}=k_{U^c_i}=k_{D^c_i}= \{-2,0,+2\}, \qquad k_{H_{u,d}}=0.\eeq 
This is similar to the $\SL(2,\mathbb{Z})$ model of eq.\eq{modelfull}~\cite{Feruglio:2023uof}, except that 
the largest weight is now 2, in absolute value.
So these models might be derivable in simpler string compactifications. 
The gauge anomalies of eq.~(\ref{sys:ano1}) cancel when summing over generations.
The relevant part of the theory is described by
\begin{eqnsystem}{sys:KWf}
K&=& Q^\dagger y^{-k_Q}Q+D^{c\dagger} y^{-k_{D^c}} D^c+ U^{c\dagger} y^{-k_{U^c}} U^c+H_d^\dagger H_d+ H_u^\dagger H_u\\
W&=& Q^T m^d D^c+Q^T m^u U^c\\
f&=& f_{0},
\end{eqnsystem}
where the quark mass matrices allowed by the modular flavor symmetry are
\begin{equation}
 m^q|_\text{can} = v_q \begin{pmatrix}
 0 & 0 & c^q_{13} \\
 0 & c^q_{22} & y \left[c^q_{23} Z_1^{(2)} + c^{\prime q}_{23} Z_2^{(2)}\right] \\
 c^q_{31} & y \left[c^q_{32} Z_1^{(2)} + c^{\prime q}_{32} Z_2^{(2)}\right] & 
 y^2 \left[c^q_{33} Z_1^{(4)} + c^{\prime q}_{33} Z_2^{(4)} + c^{\prime\prime q}_{33} Z_3^{(4)}\right]
 \end{pmatrix}\,.
\end{equation}
The structure is similar to the one in~\cite{Feruglio:2023uof}, 
with $E_4(\tau)$ and $E_6(\tau)$ replaced by
different modular forms of level 2, 
namely, $Z^{(2)}_{1,2}(\tau)$ and $Z^{(4)}_{1,2,3}(\tau)$.
The determinant of $m^q$ is real, and the existence of multiple modular forms leads to a non-vanishing CKM phase.
As in~\cite{Feruglio:2023uof} the physical quark mass matrices are obtained after making the kinetic terms in $K$ canonical,
thereby dressing the holomorphic matrices $m^q$ with the non-holomorphic factors
arising from the transformation that renders the kinetic terms canonical, 
as discussed in section~\ref{HQ}. 
This results in the powers of $y = 2\,{\rm Im}\,\tau$ 
present in the entries 23, 32 and 33.

Table~\ref{tab:data} summarizes the observed values of the quark masses, mixings and the CKM phase, renormalized at the unification scale.
All these observables can be reproduced, as the number of free parameters is large enough. 
Namely, in addition to the VEV of $\tau$, we have
10 coefficients $c^q_{ij}$ in each quark sector.

Scanning over $\tau$ and the 18 ratios of real parameters $c^q_{ij}/c^q_{13}$ ($ij\neq13$), 
we find the following sample point describing well the quark data: 
\begin{align}
 \tau &= -0.182 + 1.283\, i\,, \\[0.2cm]
 c^u &= 10^{-3} \begin{pmatrix}
 0 & 0 & 1.22 \\
 0 & 2.20 & -20.3,\,35.0 \\
 0.96 & 16.9,\,7.72 & 61.7,\,-10.1,\,-12.2
 \end{pmatrix}\,, \\[0.2cm]
 c^d &= 10^{-3} \begin{pmatrix}
 0 & 0 & 0.22 \\
 0 & 1.01 & -1.58,\,4.92 \\
 0.92 & 6.89,\,-4.42 & 1.83,\,11.4, -5.00
 \end{pmatrix}\,.
\end{align}
Assuming $\tan\beta = v_u/v_d = 5$,
the overall scales $v_q c^q_{13}$ are fixed by reproducing
$m_t = 92.97$~GeV and $m_b = 1.21$~GeV at the unification scale $\sim 2~10^{16}$~GeV. 
In this solution the coefficients $c$ are not comparable:
modular forms alone do not fully generate the quark mass hierarchies.
The normalization of modular forms is anyhow arbitrary.


\subsection{Models with quarks in doublets of $\Gamma_2$}\label{Gamma_2}
We next consider a model where flavor is non-trivially embedded into the modular group, leading to a non-abelian structure.
Specifically, we here assume that the flavor symmetry group is the full $\SL(2,\mathbb{Z})$, but
flavor is non-trivially embedded into it, such that the matter fields transform nontrivially under $\Gamma_2$.
In the lepton sector, this can help understanding the large neutrino mixing angles~\cite{1803.10391,Meloni:2023aru,2402.18547}.

In the quark sector the mixing angles are small, so a non-abelian structure can be tolerated 
in the presence of heavy vector-like quarks and potentially lead to a lower number of free parameters resulting in a higher predictive power.
As discussed in section~\ref{HQ}, our proposed solution to the strong CP problem is general enough
to work with an extended matter content.

We add an extra vector-like quark $D'_i \oplus D'^c_i$ for each right-handed down quark $D^c_i$,
and similarly an extra vector-like quark $U'_i \oplus U'^c_i$ for each right-handed up quark $U^c_i$.
Table~\ref{tab:Gamma_2_doublets_singlets_weight4} shows the assignment of $\Gamma_2$ representations and weights.
All generation triplets are assigned to a doublet plus a suitable singlet, 
$\mathbf{1}_0$ or $\mathbf{1}_1$.
The Higgs multiplets $H_{u,d}$ are assumed to be complete singlets $\mathbf{1}_0$ with vanishing weight. 
The negative weights of the SM quarks are balanced by the positive weights of the vector-like quarks, 
cancelling the QCD anomaly of eq.\eq{ano13}.
Extra heavy leptons can be added to cancel the electroweak anomalies in eq.~(\ref{sys:ano1}),
and extra singlets can be added to cancel gravitational anomalies.

\begin{table}[t]
\centering
\resizebox{1.0\textwidth}{!}{
\begin{tabular}{|c|ccc|cccc|} \hline
& \multicolumn{3}{c|}{SM quarks} & \multicolumn{4}{c|}{Extra vector-like quarks}\\
& \cellcolor{gray!10} $Q$& \cellcolor{blue!10}  $D^c$& \cellcolor{red!10}   $U^c$  &  \cellcolor{blue!10} $D'^c$ &  \cellcolor{blue!10} $D'$ & \cellcolor{red!10}  $U'^c$ & \cellcolor{red!10}  $U'$ \\
\hline
${\rm SU}(2)_L\otimes{\rm U}(1)_Y$ &  \cellcolor{gray!10} $2_{1/6}$&  \cellcolor{blue!10} $1_{1/3}$&  \cellcolor{red!10} $1_{-2/3}$&
\cellcolor{blue!10}   $1_{1/3}$& \cellcolor{blue!10} $1_{-1/3}$ &  \cellcolor{red!10}  $1_{-2/3}$
&  \cellcolor{red!10} $1_{2/3}$\\
Flavor symmetry $\Gamma_2$& \cellcolor{gray!10} ${\bf 2}\oplus{\bf 1}_0$ & \cellcolor{blue!10} ${\bf 2}\oplus{\bf1}_1$ & \cellcolor{red!10}  ${\bf 2}\oplus{\bf 1}_0$ &
 \cellcolor{blue!10} ${\bf2}\oplus{\bf 1}_0$ & \cellcolor{blue!10} ${\bf 2}\oplus{\bf 1}_1$
&  \cellcolor{red!10} ${\bf 2}\oplus{\bf 1}_0$ & \cellcolor{red!10}  ${\bf 2}\oplus{\bf 1}_0$\\
Modular weights  $k_\Phi$& \cellcolor{gray!10} $-2$& \cellcolor{blue!10} $-2$& \cellcolor{red!10} $-2$ &
\cellcolor{blue!10}  $+2$& \cellcolor{blue!10} $+2$& \cellcolor{red!10}  $+2$&  \cellcolor{red!10}  $+2$\\
\hline
\end{tabular} 
}
\caption{\em Electroweak quantum numbers, weights and $\Gamma_2$ representations of quarks in the model of section~\ref{Gamma_2}.}
\label{tab:Gamma_2_doublets_singlets_weight4}
\end{table}

The action is specified by
\begin{eqnsystem}{sys:WHQ}
K_{\rm UV}&=& 
Q^\dagger y^{-k_Q}Q+
D^{c\dagger} y^{-k_{D^c}} D^c+
D'^\dagger y^{-k_{D'}}D'  + 
D'^{c\dagger} y^{-k_{D'^c}} D'^c
+  H_d^\dagger H_d+
\\
&+&U^{c\dagger} y^{-k_{U^c}} U^c  +
U'^\dagger y^{-k_{U'}}U'   +
U'^{c\dagger}y^{-k_{U'^c}} U'^c + 
H_u^\dagger H_u, \nn
\\
W_{\rm UV}
&=& Q^T m^d D^c + 
Q^T n^d D'^c  + 
D'^T N^d D^c+ 
D'^T M^d D'^c +
\\
&+&
Q^T m^u U^c+ 
Q^T n^u U'^c+ 
U'^T N^u U^c+ 
U'^T M^u U'^c,
\nn \\ 
f_{\rm UV}&=&~f_{0},
\end{eqnsystem}
where the mass matrices $m^q$, $n^q$, $N^q$ and $M^q$ 
follow the notation of section~\ref{HQ}.
In the up sector, we have
\begin{align}
 m^u &= \mathbb{0}_{3\times3}\,,
 \qquad
 n^u = n_u \begin{pmatrix}
 1 & 0 & 0 \\
 0 & 1 & 0 \\
 0 & 0 & \alpha_u
 \end{pmatrix}\,,
 \qquad
 N^u = N_u \begin{pmatrix}
 1 & 0 & 0 \\
 0 & 1 & 0 \\
 0 & 0 & \beta_u 
 \end{pmatrix}\,, \\[0.2cm]
 M^u &= M_u \begin{pmatrix}
 -Z^{(4)}_1 + \gamma_{u1} Z^{(4)}_3 & Z^{(4)}_2 & \gamma_{u2} Z^{(4)}_1 \\
 Z^{(4)}_2 & Z^{(4)}_1 + \gamma_{u1} Z^{(4)}_3 & \gamma_{u2} Z^{(4)}_2 \\
 \gamma_{u3} Z^{(4)}_1 & \gamma_{u3} Z^{(4)}_2 & \gamma_{u0} Z^{(4)}_3
 \end{pmatrix}\,.
\end{align}
Here, $Z^{(4)}_i$, $i=1,2,3$, are the modular forms of level 2 and weight 4 
defined in eq.~\eqref{eq:N2k4}.
$(Z^{(4)}_1,Z^{(4)}_2)$ form a doublet of $\Gamma_2$, whereas 
$Z^{(4)}_3$ is an invariant singlet.
In the down sector, we have
\begin{align}
 m^d &= \mathbb{0}_{3\times3}\,,
 \qquad
 n^d = n_d \begin{pmatrix}
 1 & 0 & 0 \\
 0 & 1 & 0 \\
 0 & 0 & \alpha_d
 \end{pmatrix}\,,
 \qquad
 N^d = N_d \begin{pmatrix}
 1 & 0 & 0 \\
 0 & 1 & 0 \\
 0 & 0 & \beta_d 
 \end{pmatrix}\,, \\[0.2cm]
 M^d &= M_d \begin{pmatrix}
 -Z^{(4)}_1 + \gamma_{d1} Z^{(4)}_3 & Z^{(4)}_2 & \gamma_{d2} Z^{(4)}_1 \\
 Z^{(4)}_2 & Z^{(4)}_1 + \gamma_{d1} Z^{(4)}_3 & \gamma_{d2} Z^{(4)}_2 \\
 \gamma_{d3} Z^{(4)}_2 & -\gamma_{d3} Z^{(4)}_1 & 0
 \end{pmatrix}\,.
\end{align}
The $n_q $ have weak-scale value, as they arise from the Higgs VEV $v_q$.
The $M_q$ and $N_q$ can be arbitrarily large, up to the string scale.
To avoid new states around the weak scale we assume $M_q, N_q \gg v$. 
So, in what follows, we will neglect $n_q/M_q \ll 1$.

The determinants of the full quark matrices 
$\mathcal{M}^u$ and $\mathcal{M}^d$, and of their product, are real:
\begin{equation}
 \det\mathcal{M}^q = - n_q^3 N_q^3 \alpha_q \beta_q 
 \qquad \text{so} \qquad 
 \det(\mathcal{M}^u\mathcal{M}^d) = n_u^3 n_d^3 N_u^3 N_d^3 \alpha_u \alpha_d \beta_u \beta_d\,.
 \label{eq:Gamma_2_detfullM}
\end{equation}
Since the low-energy theory has a modular anomaly, the determinant of the holomorphic part 
of the light quark mass matrix 
$m^q_\mathrm{hol} = -n^q \left(M^q\right)^{-1} N^q$ depends on $\tau$:
\begin{equation}
 \det m^q_\mathrm{hol} = - \frac{n_q^3 N_q^3 \alpha_q \beta_q}{\det M^q(\tau)}\,.
\end{equation}
The strong CP problem is solved if the real determinant of 
$\mathcal{M}^u\mathcal{M}^d$ in eq.~\eqref{eq:Gamma_2_detfullM} is positive.
In the low-energy theory, this solution is reproduced by a modification of the low-energy gauge kinetic function of the color $\SU(3)_c$ gauge group, which reads
\begin{equation}
 f_{\rm IR}=f_0-\frac{1}{8\pi^2}\ln\det M^u M^d.
\end{equation}
The determinant
\begin{equation}
 \det M^d = M_d^3 \gamma_{d2}\gamma_{d3} \left({Z^{(4)}_2}^3 - 3 {Z^{(4)}_1}^2 Z^{(4)}_2\right)
\end{equation}
is a modular form of weight 12, transforming in the singlet $\mathbf{1}_1$ of $\Gamma_2$,
that vanishes at $\tau=i$.
So $m^d_\mathrm{hol}$  develops a pole at this point, and the low-energy description fails if $\tau$ is too close to this point. 
In the up sector,  the dependence of $\det M^u$ 
on the parameters $\gamma_{ui}$ does not factorise, so it might vanish at different values of $\tau$.

\begin{table}
\centering
\renewcommand{\arraystretch}{1.25}
\begin{tabular}[t]{|l|c|l|c|l|c|}
\hline
\multicolumn{6}{|c|}{Input parameters} \\
\hline
\multicolumn{2}{|c|}{Modulus} & \multicolumn{2}{c|}{Up sector} & \multicolumn{2}{c|}{Down sector} \\
\hline
$\Re\tau$ & $-0.4347$ & $\alpha_u$ & $13.71$ & $\alpha_d$ & $13.30$ \\
$\Im\tau$ & $1.646$ & $\beta_u$ & $-0.0723$ & $\beta_d$ & $-4.681$ \\
 & & $\gamma_{u0}$ & $0.1087$ & & \\
 & & $\gamma_{u1}$ & $-0.2442$ & $\gamma_{d1}$ & $1.105$ \\
 & & $\gamma_{u2}$ & $-0.0216$ & $\gamma_{d2}$ & $-0.2165$ \\
 & & $\gamma_{u3}$ & $-62.38$ & $\gamma_{d3}$ & $-0.0460$ \\
 & & $N_u/M_u$ & $1.515$ & $N_d/M_d$ & $0.6220$ \\
\cline{3-6}
 & & $n_u$~[GeV] & $24.82$ & $n_d$~[GeV] & $0.3411$ \\
\hline
\end{tabular}
\hspace{0.2cm}
\begin{tabular}[t]{|c|c|}
\hline
\multicolumn{2}{|c|}{Output values} \\
\hline
$m_u/m_c$ & $1.46 \times 10^{-3}$ \\
$m_c/m_t$ & $2.68 \times 10^{-3}$ \\
$m_d/m_s$ & $4.99 \times 10^{-2}$ \\
$m_s/m_b$ & $1.42 \times 10^{-2}$ \\
$\sin^2\theta_{12}$ & $5.06 \times 10^{-2}$ \\
$\sin^2\theta_{13}$ & $1.03 \times 10^{-5}$ \\
$\sin^2\theta_{23}$ & $1.25 \times 10^{-3}$ \\
$\delta/\pi$ & $0.391$\\
\hline
$\chi^2$ & 0.47 \\
\hline
\end{tabular}
\caption{\em Best-fit values of the input parameters 
and resulting dimensionless observables 
in the model of section~\ref{Gamma_2}.}
\label{tab:Gamma_2_doublets_singlets_weight4_bfp}
\end{table}

The light quark mass matrices  depend also on the non-holomorphic factors, see section~\ref{HQ}:
\begin{equation}
 m^q_\text{IR}|_\text{can} = - y^{-4} n^q \left(M^{q}\right)^{-1} N^q
 \left[\mathbb{1} + y^{-6} N^{q\dagger} \left(M^q M^{q\dagger}\right)^{-1} N^q\right]^{-1/2}\,.
\end{equation}
%
In addition to $\tau$, the up sector contains 7 dimensionless parameters: $\alpha_u$, $\beta_u$, $\gamma_{ui}$, $i=0,1,2,3$, and $N_u/M_u$, 
whereas the down sector is characterised by 6 dimensionless parameters:
$\alpha_d$, $\beta_d$, $\gamma_{d1,2,3}$ and $N_d/M_d$. 
A numerical scan over these parameters shows that the model can fit the data in table~\ref{tab:data}
for the values of the 
parameters shown in table~\ref{tab:Gamma_2_doublets_singlets_weight4_bfp}.
Some parameters have mildly hierarchical values.
The mass matrices have the natural hierarchical onion structure.
All heavier generations components dominantly lie in singlets, except for $t_R$ which lies in a flavor doublet.

The existence of this fit is non-trivial, as non-abelian mass matrices can imply quark masses and mixings qualitatively different from the observed ones,
even when the number of free parameters is larger than the number of data.
Fitting data can be impossible assuming different assignments of the singlet representations $\mb{1}_{0,1}$.

The overall scale of the  heavy quark mass is
not fixed and can be taken arbitrarily large, for example around the unification scale
in order not to spoil the unification of gauge couplings, and not to affect the running of quark masses and mixings.


\section{Models with modular symmetry at level $N=3$}\label{N=3}
We next consider the level $N=3$. The infinite-dimensional principal congruence subgroup $\Gamma(3)$ 
is generated by three elements that can be chosen as
\be\nn
G_1=
\left(
\begin{array}{cc}
1&3\\
0&1
\end{array}
\right)=T^3,\qquad
G_2=
\left(
\begin{array}{cc}
-8&3\\
-3&1
\end{array}
\right)=(T^3S)^2S^2 ,\qquad
G_3=
\left(
\begin{array}{cc}
4&-3\\
3&-2
\end{array}
\right)=(T^2S)^3.
\ee
The quotient group $\Gamma_3=\SL(2,\mathbb{Z})/\Gamma(3)$ is isomorphic to $T'$,
the double cover of the tetrahedral group $A_4$, which is a geometrically intuitive discrete sub-group of SO(3).
$T'$ is the corresponding  discrete sub-group of SU(2), which is the double cover of SO(3).
$T'$ has dimension 24 and is generated by two elements ${\tS}$ and ${\tT}$ satisfying:\footnote{An equivalent presentation is: $\tS^2=- \One$ and
$(\tS\tT)^3=\tT^3=\One$.}
\beq\label{eq:STcal}
\tS^2=(\tS\tT)^3=- \One ,\qquad \tT^3=\One.
\eeq
Its irreducible representations are three singlets ${\bf 1}_i$, three doublets ${\bf 2}_i$ related to spinorial representations
and one triplet $\bf 3$. The index $i$ conveniently runs over $\{0,1,2\}$.
Explicit expressions of the generators ${\tS}$ and ${\tT}$ in the various representations, tensor products and 
Clebsch-Gordan coefficients are given in appendix \ref{g3}.

\medskip

The space ${\cal M}_k(\Gamma(3))$ of the modular forms of level 3 and weight $k$ has dimension $k+1$. 
So, at weight $k=1$,
there are two linearly independent modular forms: $Z^{(1)}_{a}$ with $a=\{1,2\}$.
They can be built out of the Dedekind eta function $\eta(\tau)$~\cite{1907.01488}:
\be
Z^{(1)}_{1}=\sqrt{2}~\frac{\eta^3(3\tau)}{\eta(\tau)},
\qquad
Z^{(1)}_{2}=\frac{\eta^3(3\tau)}{\eta(\tau)}+\frac{\eta^3(\tau/3)}{3~\eta(\tau)}.
\label{eq:N3k1}
\ee
From the transformation properties of $\eta(\tau)$, it follows that under the group $\Gamma(3)$ they transform as in eq.~(\ref{modtrans}) with $k=1$.
Higher weights modular forms are homogeneous polynomials in the variables $Z^{(1)}_{1,2}$. 
For instance, the 3 linearly independent modular forms at weight 2 can be built as\footnote{They are related to the three forms $Y_i(\tau)$ of~\cite{Feruglio:2017spp} as $(Y_1(\tau),Y_2(\tau),Y_3(\tau))=-9 (Z_1(\tau),Z_3(\tau),Z_2(\tau))$.}
\be
Z^{(2)}_a =\left\{ -(Z^{(1)}_{2})^2,(Z^{(1)}_{1})^2,\sqrt{2}~Z^{(1)}_{1}Z^{(1)}_{2}  \right\}.
\label{eq:N3k2}
\ee
Under the full $\SL(2,\mathbb{Z})$, the two forms $Z^{(1)}_{1,2}$ fill a doublet,  
and $Z^{(2)}_{1,2,3}$ fill a triplet, transforming as:
\begin{eqnsystem}{sys:Z12}
Z^{(1)}_{}(\gamma\tau)=&~(c\tau+d)~\rho_{{\bf 2}_2}(\gamma)Z^{(1)}_{}(\tau)\,,\\
Z^{(2)}_{}(\gamma\tau)=&~(c\tau+d)^2\rho_{{\bf 3}}(\gamma)Z^{(2)}_{}(\tau)\,,
\end{eqnsystem}
where the matrixes
$\rho_{{\bf 2}_2}$ and $\rho_{{\bf 3}}$ are the unitary representations ${\bf 2}_2$ and ${\bf 3}$ of $\Gamma_3 \cong T'$, see appendix~\ref{g3}.

\subsection{A model with $\Gamma(3)$ modular symmetry}\label{Gamma(3)}
We here assume that $\Gamma(3)$ is the flavor symmetry group, and propose a supersymmetric
model that only
includes the three  generations of SM quarks (i.e.~no heavy quarks),
$Q_{1,2,3}$, $D^c_{1,2,3}$ and $U^c_{1,2,3}$, and the MSSM Higgs doublets $H_{u,d}$.
We assume the minimal weights
\beq \label{eq:model11}
k_{Q_i}=k_{U^c_i}=k_{D^c_i}= \{-1,0,+1\},\qquad k_{H_{u,d}}=0.\eeq 
This is similar to the $\SL(2,\mathbb{Z})$ model of eq.\eq{modelfull}~\cite{Feruglio:2023uof}
and to the $\Gamma(2)$ model of section~\ref{Gamma(2)}, except that 
the largest weight is now 1, in absolute value.
So these models might be derivable in simpler string compactifications. 
The gauge anomalies of eq.~(\ref{sys:ano1}) cancel when summing over generations.
The relevant part of the theory is described by a super-potential $W$, a Kahler metric $K$ and a gauge kinetic function as in eq.~(\ref{sys:KWf}),
with the only difference that the  quark mass matrices allowed by the flavor symmetry now are
\begin{equation}
 m^q|_\text{can} = v_q \begin{pmatrix}
 0 & 0 & c^q_{13} \\
 0 & c^q_{22} & \sqrt{y}\left[c^q_{23} Z_1^{(1)} + c^{\prime q}_{23} Z_2^{(1)}\right] \\
 c^q_{31} & \sqrt{y} \left[c^q_{32} Z_1^{(1)} + c^{\prime q}_{32} Z_2^{(1)}\right] & 
 y \left[c^q_{33} Z_1^{(2)} + c^{\prime q}_{33} Z_2^{(2)} + c^{\prime\prime q}_{33} Z_3^{(2)}\right]
 \end{pmatrix}\,,
\end{equation}
where the level 3 modular forms $Z_{1,2}^{(1)}$ and $Z_{1,2,3}^{(2)}$ are given in eq.s~\eqref{eq:N3k1} and \eqref{eq:N3k2}, respectively.
The determinant is real, and the existence of multiple modular forms leads to a non-vanishing CKM phase.
The physical quark mass matrices are obtained after making the kinetic terms in $K$ canonical,
thereby dressing the holomorphic matrices $m^q$ with the non-holomorphic factors
arising from the transformation that renders the kinetic terms canonical, 
see section~\ref{HQ}. 
All observed quark masses and mixings can be reproduced, as the number of free parameters is large enough.
As in the model of section~\ref{Gamma(2)}, we have 10 Lagrangian 
parameters per sector, in addition to $\tau$.
Performing numerical minimization, we find that the following point
can successfully describe the quark data:
\begin{align}
 \tau &= 0.351 + 2.092\, i\,, \\[0.2cm]
 c^u &= 5~ 10^{-3} \begin{pmatrix}
 0 & 0 & 1.12 \\
 0 & 0.13 & 9.53,\,150 \\
 0.11 & -9.59,\,0.51 & 8.56,\,0.48,\,11.6
 \end{pmatrix}\,, \\[0.2cm]
 c^d &= 5~ 10^{-3} \begin{pmatrix}
 0 & 0 & 0.08 \\
 0 & 1.24 & 0.05,\,8.12 \\
 0.017 & -2.58,\,0.046 & -0.018,\,0.18, -0.020
 \end{pmatrix}\,.
\end{align}
Assuming $\tan\beta = v_u/v_d = 5$,
the overall scales $v_q c^q_{13}$ are fixed by reproducing
$m_t = 92.97$~GeV and $m_b = 1.21$~GeV at the unification scale of $\sim 2~10^{16}$~GeV.

\subsection{Models with quarks in triplets of $\Gamma_3$}\label{sec:3Gamma3}
We next consider supersymmetric  models where the flavor symmetry group is the full $\SL(2,\mathbb{Z})$, but
the three generations are non-trivially embedded into it, such that the quarks fill non-abelian representations of $\Gamma_3$.
Given that 3 generations exist, the nicest option is to embed all quarks in the ${\mb 3}$ representation,
possibly with the minimal $\pm1$ weights.
The Higgs multiplets $H_{u,d}$ are assumed to be singlets with vanishing modular weight. 

However,  fitting quark masses and mixings requires some amount of extra heavy quarks.

\begin{table}[t]
\centering
\begin{tabular}{|c|ccc|cccc|} \hline
& \multicolumn{3}{c|}{SM quarks} & \multicolumn{4}{c|}{Extra vector-like quarks}\\
& \cellcolor{gray!10} $Q$& \cellcolor{blue!10}  $D^c$& \cellcolor{red!10}   $U^c$  &  \cellcolor{blue!10} $D'^c$ &  \cellcolor{blue!10} $D'$ & \cellcolor{red!10}  $U'^c$ & \cellcolor{red!10}  $U'$ \\
\hline
${\rm SU}(2)_L\otimes{\rm U}(1)_Y$ &  \cellcolor{gray!10} $2_{1/6}$&  \cellcolor{blue!10} $1_{1/3}$&  \cellcolor{red!10} $1_{-2/3}$&
\cellcolor{blue!10}   $1_{1/3}$& \cellcolor{blue!10} $1_{-1/3}$ &  \cellcolor{red!10}  $1_{-2/3}$
&  \cellcolor{red!10} $1_{2/3}$\\
Flavor symmetry $\Gamma_3$& \cellcolor{gray!10} ${\bf 3}$ & \cellcolor{blue!10} ${\bf 3}$& \cellcolor{red!10} ${\bf 3}$&
$ \cellcolor{blue!10} {\bf 3}$& \cellcolor{blue!10} ${\bf 3}$
&  \cellcolor{red!10} ${\bf 3}$& \cellcolor{red!10}  ${\bf 3}$\\
Modular weights  $k_\Phi$& \cellcolor{gray!10} $-1$& \cellcolor{blue!10} $\pm 1$& \cellcolor{red!10} $\pm 1$ &
\cellcolor{blue!10}  +1& \cellcolor{blue!10} $\mp 1$& \cellcolor{red!10}  +1&  \cellcolor{red!10}  $\mp1$\\
\hline
\end{tabular} 
\caption{\label{tab:Gamma3A}\em Electroweak and $\Gamma_3$ representations of quarks and their weights in the model of section~\ref{sec:3Gamma3}.}
\end{table}

\medskip

Table~\ref{tab:Gamma3A} presents the simplest non-trivial model.
It adds an extra vector-like quark $D'_i \oplus D'^c_i$ for each right-handed down quark $D^c_i$,
and an extra vector-like quark $U'_i \oplus U'^c_i$ for each right-handed up quark $U^c_i$.
The modular/QCD anomaly of eq.\eq{ano13} vanishes.
Leptons are left unspecified, and must be assigned such that the electroweak anomalies in eq.~(\ref{sys:ano1}) cancel.
The arbitrary $\pm$ signs correspond to reordering $D^c$ with $D'^c$
and $U^c$ with $U'^c$, giving the same model. We fix signs such that $Q,D^c,U^c$ have $-1$ weight and,
in the notation of section~\ref{HQ},  
$M$ (rather than $N$) has a non-trivial $\tau$ dependence.
This model predicts the following mass matrices
\beq
m^q=\mathbb{0}_{3\times3}\,,
\qquad
n^q=n_q\left(
\begin{array}{ccc}
1&0&0\\
0&0&1\\
0&1&0
\end{array}
\right),
\qquad
N^q= N_q\left(
\begin{array}{ccc}
1&0&0\\
0&0&1\\
0&1&0
\end{array}
\right),\eeq
\beq
M^q=M_q\left(
\begin{array}{ccc}
 2 c^q_S Z^{(2)}_1 &  (c^q_A-c^q_S)Z^{(2)}_3 & - (c^q_A+c^q_S) Z^{(2)}_2 \\
 -(c^q_A+c^q_S) Z^{(2)}_3 & 2 c^q_S Z^{(2)}_2 &  (c^q_A-c^q_S)Z^{(2)}_1 \\
 (c^q_A-c^q_S) Z^{(2)}_2 & - (c^q_A+c^q_S)Z^{(2)}_1 & 2 c^q_S Z^{(2)}_3 \\
\end{array}
\right),
\eeq
where $n_{u,d} = y_{u,d} v_{u,d}$, $N_{u,d}$ and $M_{u,d} c^{u,d}_{A,S}$ are 8 free real parameters,
and $Z_{1,2,3}^{(2)}(\tau)$ are modular forms.
The light quark mass matrix 
\begin{equation}
 m^q_\text{IR}|_\text{can} = - y^{-2} n^q \left(M^{q}\right)^{-1} N^q
 \left[\mathbb{1} + y^{-3} N^{q\dagger} \left(M^q M^{q\dagger}\right)^{-1} N^q\right]^{-1/2}
\end{equation}
only depends on 8 combinations of parameters, for example
$\tau$, $n_q$, $c_A^q/c_S^q$, $N_q/M_q$.
A numerical study finds that the 10 quark masses and mixings cannot be reproduced.
Some predictions are qualitatively wrong, as the mass matrices have a non-abelian structure that in no limit reduces to 
an onion-like structure that leads to hierarchical quark masses with mildly small angles $\theta_{ij}\sim \sqrt{m_i/m_j}$, as observed.



The same problem arises assuming heavy $Q'\oplus Q'^c$  rather than $U'\oplus U'^c\oplus D'\oplus D'^c$, as this choice results in equivalent models.
A model with both right-handed and left-handed heavy quarks again has the same problem, as
a scan of possible assignments of $\pm1$ weights
shows that two matrices with non-trivial $\tau$-dependence (and thereby multiple free parameters)
are never simultaneously relevant in the limit $M,N\gg m,n$).
Extra free parameters are needed to fit the data.
Adding more heavy quarks such as $D''\oplus D''^c$ adds more free parameters;
however the effective low-energy mass matrix still keeps the same form, depending again on 8 effective parameters only.

To reproduce all observed quark masses and mixings one needs to drop the assumption that K\"ahler kinetic terms are minimal, see e.g.~\cite{1909.06910,2101.08718}.
Generic kinetic matrices for $U^c$, $D^c$ and $Q$ allow of course to fit data.

\medskip

Alternatively, extra free parameters can be obtained keeping minimal K\"ahler terms 
and moving to non-minimal models,
increasing the modular weights and/or by assigning some quarks to smaller representations
of $\Gamma_3$ such as ${\bf 1}_0\oplus {\bf 1}_1\oplus {\bf 1}_2$ or ${\bf 2}_i \oplus {\bf 1}_j$, with $i,j=0,1,2$.



\section{Conclusions}
Solutions to the strong CP problem based on spontaneous CP breaking
can be achieved in a large class of models, extending considerably the original Nelson-Barr realizations. In models of this class, Yukawa couplings are dynamical variables depending
on a set of complex scalar fields $z$, either elementary or composite, whose VEVs determine the fermion mass spectrum and 
spontaneously break CP. In this way, the flavor puzzle and the strong 
CP problem are solved at once. 
In~\cite{Feruglio:2023uof} we have shown that, unlike typical Nelson-Barr realizations, these solutions require no extra heavy quarks, no fine-tuning among different scales
and no more than a single complex scalar field, the modulus.
 
In the present work, we identify a general set of conditions 
characterising solutions to the strong CP problem based on 
the idea that CP is part of a local flavor symmetry,
with no QCD anomalies and spontaneously broken by fields $z$ different from the SM Higgs.
The Yukawa couplings are required to be homogeneous polynomials in the variables $z$. Moreover, the degrees
of these polynomials should be chosen so that the determinant of the quark mass matrix is 
a homogeneous polynomial of degree zero, that is a real constant. 
A non-vanishing CKM phase of the required size can arise from 
the non-trivial dependence of the Yukawa couplings on $z$.
 
From the physical point of view, these conditions can be met in a supersymmetric theory by asking the invariance of the theory under local
flavor transformations. 
The holomorphic dependence of the Yukawa couplings on $z$ is guaranteed by supersymmetry. 
The polynomial character of Yukawa couplings is enforced by assigning
matter fields a charge or a weight, depending on whether
the flavor group is U(1) or a discrete modular group, respectively.
We have shown that, when the charge/weight assignment secures the cancellation of mixed flavor/QCD anomalies, 
the determinant of the quark Yukawa couplings is indeed a real constant.

\smallskip 
 
Within this setup, only a limited number of independent patterns of Yukawa couplings is allowed,
and we have classified the non-singular ones arising when no extra quarks are included.
The general case can also involve heavy vector-like quarks. In theories containing heavy $\SU(2)_L$ singlet vector-like quarks,
we have derived the most general low-energy effective theory, showing how the cancellation of flavor-gauge anomalies is achieved thanks to an infrared gauge kinetic function with a non-trivial dependence on the fields $z$.
Nelson-Barr models represent a particular case of this class of models, where the CKM phase arises entirely from a wave function renormalization of the light quarks.

\smallskip

Models based on a U(1) flavor symmetry require that it is spontaneously broken by multiple scalar (super)fields that acquire different phases,
such that CP too gets spontaneously broken.
We do not construct specific models based on U(1), because all such ingredients are naturally present in models based on modular invariance $\text{SL}(2,\mathbb{Z})$.
Indeed, modular invariance is automatically broken by the VEV of a modulus, and modular forms effectively behave as multiple scalars with different CP phases.
Despite its unusual mathematics, modular invariance has a simple physical origin
in compactifications of string theory that provide a complex structure, allowing chirality and a non-trivial CP structure.

In the final sections of this work, we provided new concrete realizations of the general framework, solving the strong CP problem using modular invariance in novel ways: 
\begin{itemize}
\item We proposed two models based on the principal congruence sub-groups $\Gamma(2)$ and $\Gamma(3)$ of the modular group.
Such models employ no extra heavy vector-like quarks, and the largest weight in absolute value assigned to the SM quarks is $2$ in the $\Gamma(2)$ model, and $1$ in the $\Gamma(3)$ model.
These models employ lower weights (as typical in string compactifications)
than the models based on the full modular invariance $\Gamma(1)= \text{SL}(2,\mathbb{Z})$ presented in \cite{Feruglio:2023uof}.
In these two examples all quarks still belong to singlet representations.

\item We proposed models where flavor is non-trivially embedded within the modular group, as motivated by string compactifications.
In a model presented in  section \ref{Gamma_2} all 3-generation families of quarks fill non-abelian representations $\bf{2}\oplus {\bf1}_{0,1}$ of 
 the finite modular group $\Gamma_2\cong S_3$.
 The largest weight is $2$ in absolute value, and heavy vector-like quarks are needed to fit all quark masses and mixings.
The model has $13$ dimensionless free parameters in addition to the complex modulus $\tau$.
It provides the first non-abelian modular solution to the strong CP problem. 
Furthermore, in section~\ref{sec:3Gamma3} we find that the simplest model, 
where all quarks have weights $\pm1$ and fill the {\bf 3} representation of $\Gamma_3\cong T'$, 
may fit quark masses and mixings only by abandoning the assumption of a minimal K\"ahler term.
\end{itemize}
A common issue of all proposed solutions to the strong CP problem is the
potentially dangerous quantum corrections to $\bar\theta$. 
Even when $\bar\theta$ vanishes at the tree level,
$\bar\theta$ is required to remain sufficiently small after all corrections to the
leading order result have been included.
For supersymmetric realizations,
these corrections have been discussed in the literature \cite{hep-ph/9303296,Hiller:2001qg}
and summarized in~\cite{Feruglio:2023uof}.
Two main sources of additional contributions to $\bar\theta$ should be kept under
control.
The first one consists of all gauge singlets, other than the ones we have explicitly included, which can affect  $\bar\theta$
through their CP-violating VEVs.
These singlets are generically expected in realistic
cases, such as those arising from string-theory compactifications. A sufficient
condition to achieve $\bar\theta=0$ in the modular-invariant realization
is to assume that only $\tau$ acquires a CP-violating VEV. Less stringent
conditions have been discussed~\cite{Feruglio:2023uof}.
The second one is related to the specific mechanism of supersymmetry breaking. A
favorable scenario that minimizes
this second set of corrections occurs when supersymmetry breaking is gauge-mediated~\cite{hep-ph/9801271} or anomaly-mediated~\cite{Randall:1998uk} at energies below the flavor mass
scale.
In such a case, the renormalization group and threshold corrections due to
supersymmetry breaking have the same
flavor and CP structure as the SM corrections, and do not represent a threat to
our solution.

In this paper, we have chosen to focus on models invariant under rigid
supersymmetry. All the new features exhibited
here can be implemented straightforwardly in supergravity. Modular-invariant
level-1 supergravity theories
have been discussed in~\cite{Feruglio:2023uof}. The extension to the higher levels
along the
lines described in the present work, does not present either new features or more
difficulties.


\small
\paragraph{Acknowledgments.}
F.F., M.P. and A.T. are grateful to the Mainz Institute for Theoretical Physics (MITP) of the Cluster of Excellence PRISMA+ (Project ID 390831469), 
for its hospitality and partial support during the completion of this work.

\appendix\small

\section{Level $N=2$}\label{g2}

\subsection{Group properties of $\Gamma_2\cong S_3$}

The group $\Gamma_2$ is isomorphic to the group $S_3$
describing permutations of 3 elements or, in geometric terms, the symmetries of the equilateral triangle.
It has order 6 and is generated by two elements ${\tS}$ and ${\tT}$
satisfying eq.\eq{gen_s3}. 
Its irreducible representations are two singlets ${\bf 1}_0$ and ${\bf 1}_1$,
and one doublet~${\bf 2}$. Their tensor products are
\beq
{\bf1}_1\otimes {\bf 1}_1= {\bf 1}_0,\qquad 
{\bf1}_1\otimes\bf{2}=\bf{2},\qquad
\bf{2}\otimes\bf{2}={\bf1}_0\oplus{\bf1}_1\oplus\bf{2}.
\eeq
We work in the basis given in table~\ref{A.1},
\begin{table}[t]
\centering
\begin{tabular}{|c|cc|c|} 
\hline
\cellcolor{gray!10} $S_3$ & ${\bf 1}_0$ & ${\bf 1}_1$ & ${\bf 2}$ \\
\hline
$\tS$ & $1$ & $-1$ & $\tS_{\bf 2}$ \\
$\tT$ & $1$ & $-1$ & $\tT_{\bf 2}$ \\
\hline
\end{tabular} 
\hspace{1cm}
\begin{tabular}{|c|ccc|ccc|c|} 
\hline
\cellcolor{gray!10} $T'$ &${\bf 1}_0$&${\bf 1}_1$&${\bf 1}_2$&${\bf 2}_0$&${\bf 2}_1$&${\bf 2}_2$&${\bf 3}$\\
\hline
$\tS$&$1$&$1$&$1$&$\tS_{{\bf 2}_0}$&$\tS_{{\bf 2}_1}$&$\tS_{{\bf 2}_2}$&$\tS_{\bf 3}$\\
$\tT$&$1$&$\omega$&$\omega^2$&$\tT_{{\bf 2}_0}$&$\tT_{{\bf 2}_1}$&$\tT_{{\bf 2}_2}$&$\tT_{\bf 3}$\\
\hline
\end{tabular} 
\caption{\em Representations of the non-abelian finite groups $\Gamma_2 \cong S_3$ and $\Gamma_3 \cong T'$.}
\label{A.1}
\end{table}
where the $S_3$ generators in the doublet representation are given by
\begin{equation}
\label{genersss}
\tS_{\mathbf{2}}=\frac{1}{2}\begin{pmatrix}
-1&-\sqrt{3}\\
-\sqrt{3}&1
\end{pmatrix}\,,
\qquad
\tT_{\mathbf{2}}=\begin{pmatrix}
1&0\\
0&-1
\end{pmatrix}\\ \,.
\end{equation}
In this basis, the Clebsch-Gordan coefficients of the tensor products are
\begin{align}
\label{g2CB}
(\gamma_{{\bf 1}_1}\otimes \beta_{\bf{2}})_{\bf{2}}=&~(-\gamma\,\beta_2, \gamma\beta_1 )\,,\nn \\
(\alpha_{\bf{2}}\otimes \beta_{\bf{2}})_{{\bf1}_0}=&~\alpha_1\beta_1+\alpha_2\beta_2\,,\nn \\
(\alpha_{\bf{2}}\otimes \beta_{\bf{2}})_{{\bf1}_1}= &~\alpha_1\beta_2-\alpha_2\beta_1\,,\nn \\
(\alpha_{\bf{2}}\otimes \beta_{\bf{2}})_{\bf{2}}=&~( \alpha_2\beta_2-\alpha_1\beta_1, \alpha_1\beta_2+\alpha_2\beta_1)\,.\nn
\end{align}

\subsection{Modular forms of $\Gamma(2)$}

The level $N=2$ modular forms of lowest weight transform as a doublet and can be expanded in $q\equiv e^{2\pi i\tau}$ as
\begin{equation}
\label{qexp}
\begin{pmatrix}
Z_1(\tau)\\ Z_2(\tau)
\end{pmatrix}_\mathbf{2}=\begin{pmatrix}
   1 + 24  q + 24  q^2 + 96  q^3 + 24  q^4 + ... 
\\
   8 \sqrt{3}q^{1/2}(1 + 4  q + 6  q^2 + 8  q^3 +...)
\end{pmatrix}\,.
\end{equation}
For this doublet we use a different arbitrary normalization compared to \cite{1803.10391} and \cite{Meloni:2023aru}. 
See~\cite{deMedeirosVarzielas:2023crv, Petcov:2023fwh} for recent discussions on the normalization of the modular forms.

\section{Level $N=3$}\label{g3}

\subsection{Group properties of $\Gamma_3\cong T'$}\label{mod3}

The group $\Gamma_3$ is isomorphic to the double tetrahedral group $T'$. It has dimension 24 and is generated by two elements ${\tS}$ and ${\tT}$
satisfying eq.\eq{STcal}. 
Its irreducible representations are three singlets ${\bf 1}_i$, three doublets ${\bf 2}_i$ with $i=\{0,1,2\}$ and one triplet
$\bf 3$, with tensor products:
\begin{align}
{\bf 2}_i\otimes {\bf 2}_j=&~{\bf 3}\oplus{\bf 1}_{i+j~{\rm mod}~3}\,,\nn\\
{\bf 2}_i\otimes {\bf 3}=&~{\bf 2}_0\oplus {\bf 2}_1\oplus {\bf 2}_2\,,\nn\\
{\bf 3}\otimes {\bf 3}=&~{\bf 3}_S\oplus{\bf 3}_A\oplus {\bf 1}_0 \oplus {\bf 1}_1 \oplus {\bf 1}_2\,.\nn
\end{align}
We adopt the basis displayed in table \ref{A.1}, where the doublet matrices are
\beq
\tS_{{\bf 2}_i}=\frac{i}{\sqrt{3}}
\left(
\begin{array}{cc}
1&-\sqrt{2}\\
-\sqrt{2}&-1
\end{array}
\right) ,\quad
\tT_{{\bf 2}_0}=
\left(
\begin{array}{cc}
\omega^2&0\\
0&\omega
\end{array}
\right),\quad
\tT_{{\bf 2}_1}=
\left(
\begin{array}{cc}
1&0\\
0&\omega^2
\end{array}
\right), \quad
\tT_{{\bf 2}_2}=
\left(
\begin{array}{cc}
\omega&0\\
0&1
\end{array}
\right)
\eeq
and the triplet matrices are
\beq\tS_{{\bf 3}}=\frac{1}{3}
\left(
\begin{array}{ccc}
-1&2&2\\
2&-1&2\\
2&2&-1
\end{array}
\right),\qquad
\tT_{{\bf 3}}=
\left(
\begin{array}{ccc}
1&0&0\\
0&\omega^2&0\\
0&0&\omega
\end{array}
\right).
\eeq
In this basis, the Clebsch-Gordan coefficients of tensor products are:
\begin{align}
(\alpha_{{\bf 2}_i}\otimes \beta_{{\bf 2}_j})_{{\bf 1}_{i+j}}=&~(\alpha_1\beta_2-\alpha_2\beta_1)\,,\nn\\
(\alpha_{{\bf 2}_i}\otimes \beta_{{\bf 2}_{3-i}})_{{\bf 3}}=&~\left(\frac{1}{\sqrt{2}}(\alpha_1\beta_2+\alpha_2\beta_1),-\alpha_2\beta_2,\alpha_1\beta_1\right)\,,\nn\\
(\alpha_{{\bf 2}_i}\otimes \beta_{{\bf 2}_{2-i}})_{{\bf 3}}=&~\left(\alpha_1\beta_1,\frac{1}{\sqrt{2}}(\alpha_1\beta_2+\alpha_2\beta_1),-\alpha_2\beta_2\right)\,,\nn\\
(\alpha_{{\bf 2}_i}\otimes \beta_{{\bf 2}_{1-i}})_{{\bf 3}}=&~\left(-\alpha_2\beta_2,\alpha_1\beta_1,\frac{1}{\sqrt{2}}(\alpha_1\beta_2+\alpha_2\beta_1)\right)\,,\nn\\
(\alpha_{{\bf 3}}\otimes \beta_{{\bf 3}})_{{\bf 3}_S}=&~\left(2\alpha_1\beta_1-\alpha_2\beta_3-\alpha_3\beta_2,
2\alpha_3\beta_3-\alpha_1\beta_2-\alpha_2\beta_1,
2\alpha_2\beta_2-\alpha_3\beta_1-\alpha_1\beta_3
\right)\,,\nn\\
(\alpha_{{\bf 3}}\otimes \beta_{{\bf 3}})_{{\bf 3}_A}=&~\left(\alpha_2\beta_3-\alpha_3\beta_2,
\alpha_1\beta_2-\alpha_2\beta_1,
\alpha_3\beta_1-\alpha_1\beta_3
\right)\,,\nn\\
(\alpha_{{\bf 3}}\otimes \beta_{{\bf 3}})_{{\bf 1}_0}=&~\alpha_1\beta_1+\alpha_2\beta_3+\alpha_3\beta_2\,,\nn\\
(\alpha_{{\bf 3}}\otimes \beta_{{\bf 3}})_{{\bf 1}_1}=&~\alpha_2\beta_2+\alpha_3\beta_1+\alpha_1\beta_3\,,\nn\\
(\alpha_{{\bf 3}}\otimes \beta_{{\bf 3}})_{{\bf 1}_2}=&~\alpha_3\beta_3+\alpha_1\beta_2+\alpha_2\beta_1\,,\nn\\
(\alpha_{{\bf 2}_i}\otimes \beta_{{\bf 3}})_{{\bf 2}_i}=&~\left(
\alpha_1\beta_1-\sqrt2\alpha_2\beta_3,
-\alpha_2\beta_1-\sqrt2\alpha_1\beta_2
\right)\,,\nn\\
(\alpha_{{\bf 2}_i}\otimes \beta_{{\bf 3}})_{{\bf 2}_{i+1}}=&~\left(
\alpha_1\beta_3-\sqrt2\alpha_2\beta_2,
-\alpha_2\beta_3-\sqrt2\alpha_1\beta_1
\right)\,,\nn\\
(\alpha_{{\bf 2}_i}\otimes \beta_{{\bf 3}})_{{\bf 2}_{i+2}}=&~\left(
\alpha_1\beta_2-\sqrt2\alpha_2\beta_1,
-\alpha_2\beta_2-\sqrt2\alpha_1\beta_3
\right)\,.\nn
\end{align}

\subsection{Modular forms of $\Gamma(3)$}
The space ${\cal M}_k(\Gamma(3))$ of modular forms of level 3 and weight $k\ge 0$ has dimension $k+1$.
Two independent weight-1 modular forms are:
\be\nn
Z^{(1)}_{1}(\tau)=\sqrt{2}~\frac{\eta^3(3\tau)}{\eta(\tau)},\qquad
Z^{(1)}_{2}(\tau)=\frac{\eta^3(3\tau)}{\eta(\tau)}+\frac{\eta^3(\tau/3)}{3~\eta(\tau)}.
\ee
They transform as ${\bf 2}_2$ under $\Gamma_3$.
Their expansion in $q\equiv e^{2\pi i \tau}$ reads
\be\nn
Z^{(1)}_{1}=\sqrt{2} q^{1/3} \left(1+q+2 q^2+\cdots\right),\qquad
Z^{(1)}_{2}=\frac{1}{3}+2 q+2 q^3+2 q^4+\cdots .
\ee
Three independent weight-2 modular forms are:
\be\nn
Z^{(2)}=(-(Z^{(1)}_{2})^2,(Z^{(1)}_{1})^2,\sqrt{2}Z^{(1)}_{1}Z^{(1)}_{2}).
\ee
They transform as ${\bf 3}$ under $\Gamma_3$.
Their $q$-expansion  reads
\begin{align}
Z^{(2)}_{1}=&~\frac{1}{9} (-1 - 12 q - 36 q^2 - 12 q^3 - 84 q^4+\cdots ),\nn\\
Z^{(2)}_{2}=&~2q^{2/3} (1 + 2 q + 5 q^2 + 4 q^3+ \cdots), \\
Z^{(2)}_{3}=&~\frac{2}{3} q^{1/3} (1 + 7 q + 8 q^2 + 18 q^3+\cdots). \nn
\end{align}
The combination
\be\nn
({Z^{(2)}_1}^3+{Z^{(2)}_2}^3+{Z^{(2)}_3}^3-3{Z^{(2)}_1}{Z^{(2)}_2}{Z^{(2)}_3})=
\frac{1}{729} (-1 + 504 q + 16632 q^2+...)=-\frac{E_6(\tau)}{729}
\ee
has weight six and is proportional to the Eisenstein series $E_6(\tau)$. 
Out of the triplet $Z^{(2)}$ we can form the three weight-4 forms:
\beq 
Z^{(4)}_{1}=2 (Z^{(2)}_{1})^2-2 Z^{(2)}_{2}Z^{(2)}_{3}, \qquad
Z^{(4)}_{2}=2 (Z^{(2)}_{3})^2-2 Z^{(2)}_{1}Z^{(2)}_{2},\qquad
Z^{(4)}_{3}=2 (Z^{(2)}_{2})^2-2 Z^{(2)}_{3}Z^{(2)}_{1}. 
\eeq
They transform as ${\bf 3}$ under $\Gamma_3$.
The combination
\be\nn
({Z^{(4)}_1}^3+{Z^{(4)}_2}^3+{Z^{(4)}_3}^3-3{Z^{(4)}_1}{Z^{(4)}_2}{Z^{(4)}_3})=
\frac{8}{531441}(1 - 1008 q + 220752 q^2+  \cdots )=\frac{8 E_6^2(\tau)}{531441}
\ee
has weight 12 and is proportional to $E_6^2(\tau)$. 

\end{document}